\documentclass[pra,superscriptaddress,twocolumn]{revtex4-2}

\usepackage{multirow}
\usepackage{amsmath}
\usepackage{physics}
\usepackage{graphicx}
\usepackage{tikz}

\begin{document}

\title{No-cost Bell Nonlocality certification from quantum tomography and its applications in quantum-magic-resource witnessing}

\begin{abstract}
    Tomographic measurements are the standard tool for characterizing quantum states, yet they are usually regarded only as means for state reconstruction or fidelity measurement. Here, we show that the same Pauli-basis measurements (X, Y, Z) can be directly employed for the certification of nonlocality at no additional experimental cost. Our framework allows any tomographic data -- including archival datasets -- to be reinterpreted in terms of fundamental nonlocality tests. 
    We introduce a generic, constructive method to generate tailored Bell inequalities and showcase their applicability to certify the non-locality of states in realistic experimental scenarios. Recognizing the stabilizer nature of the considered operators, we analyze our inequalities in the context of witnessing quantum magic -- a crucial resource for quantum computing. Our approach requires Pauli measurements only and tests the quantum magic solely through the resources present in the state.
    Our results establish a universal standard that unifies state tomography with nonlocality certification and its application to quantum magic witnessing, thereby streamlining both fundamental studies and practical applications.
\end{abstract}

\author{Pawe{\l} Cie\'sli\'nski}
\affiliation{Institute of Theoretical Physics and Astrophysics, University of Gdańsk, 80-308 Gda\'nsk, Poland}
\affiliation{International Centre for Theory of Quantum Technologies, University of Gdańsk, 80-308 Gdańsk, Poland}

\author{Lukas Knips}
\affiliation{Max Planck Institute for Quantum Optics, 85748 Garching, Germany}
\affiliation{Faculty of Physics, Ludwig Maximilian University, 80799 Munich, Germany}
\affiliation{Munich Center for Quantum Science and Technology, 80799 Munich, Germany}

\author{Harald Weinfurter}
\affiliation{Max Planck Institute for Quantum Optics, 85748 Garching, Germany}
\affiliation{Faculty of Physics, Ludwig Maximilian University, 80799 Munich, Germany}
\affiliation{Munich Center for Quantum Science and Technology, 80799 Munich, Germany}

\author{Wies{\l}aw Laskowski}
\affiliation{Institute of Theoretical Physics and Astrophysics, University of Gdańsk, 80-308 Gda\'nsk, Poland}

\maketitle

\section{Introduction}

Bell non-locality is a defining feature of quantum mechanics and one of the most valuable resources for quantum information processing~\cite{Brunner_2014}. Besides its foundational importance, it was shown to be a source of quantum advantage in randomness generation~\cite{Pironio_2010,Colbeck_2011,Colbeck_2011_2, ArnonFriedman_2018}, device-independent cryptography~\cite{Ekert_1991,Mayers_1998, Barrett_2005, Acin_2007, Pironio_2009,Pawlowski_review_2025}, communication~\cite{Cleve_1997,Brukner_2002,Aolita_2012,Moreno_2020}, and many others. Violation of Bell inequalities was verified experimentally in a vast number of experiments~\cite{Aspect1982b, Weihs1998, Rowe2001, Pan2000, Walther2005, Hensen2015, Giustina2015, Shalm2015, Rosenfeld2017, Rauch2018, Storz2023} and is now a well-established part of modern physics and quantum science. 

Quantum state tomography can provide the full reconstruction of a multi-particle quantum state from local measurements, even if the state is highly entangled, and thus became a crucial component of many quantum mechanical experiments. During their design, quantum state tomography or a reduced set of measurements in a standard basis is usually performed to verify the validity of the prepared states. In qubit experiments, the standard measurement settings for this purpose correspond to the tensor products of Pauli matrices $\sigma_{i_1} \otimes \cdots \otimes \sigma_{i_N}$. With $N$ being the number of qubits, these tensor products form an operator basis for $2^N$-dimensional Hermitian operators. 
Note that quantum tomography can be done in many different ways, but using the Pauli matrix basis is the most convenient and common one.
Of course, one may naturally ask whether these data can give more information about the quantum state.
Here we show that, indeed, they can be used directly for the certification of quantum resources like nonlocality or magic at no additional cost. Importantly, the proposed technique allows one to directly claim the actual Bell violations, not only the existence of the Bell-type correlations (see e.g.~\cite{Zukowski_2002}).

\begin{figure}[t!]
    \centering
    \includegraphics[width=0.45\textwidth]{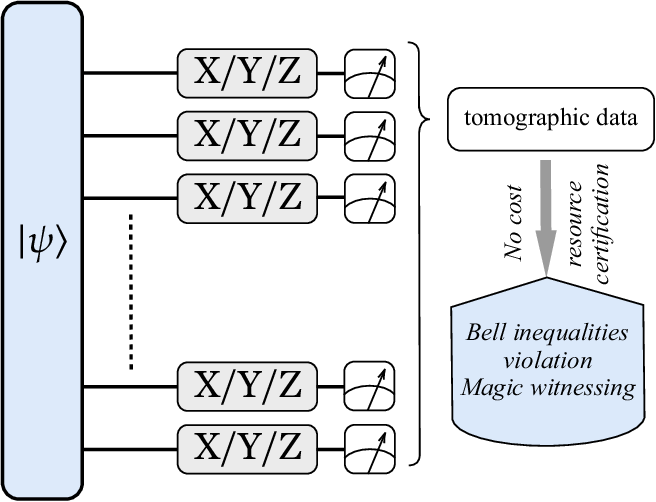}
        \caption{\textit{Can one use the tomography-type data to certify quantum resources at no additional cost?} During many quantum experiments, an experimenter has to certify the state that is being prepared in a lab. One way of doing that is by performing a set of measurements in a standard basis. For $N$-qubit systems, this is usually done by the measurements of the Pauli observables. Then the state's fidelity or full tomography can be established. Here, we explore our capabilities of exploiting this data to directly violate Bell inequalities, not only infer Bell-type correlations, and to witness quantum magic at no additional cost with a single function combining the obtained data.}
    \label{fig:placeholder}
\end{figure}

Bell inequalities using only Pauli measurements, later referred to as XYZ Bell inequalities, for the standard Bell states ($N=2$) do not exist~\cite{Howard_2015}. 
However, this limitation can be easily overcome by applying a suitable, non-stabiliser, local transformation on one of the parties, i.e. \((\openone \otimes U)|\phi\rangle\). 
In such cases, each resulting inequality reduces to the standard CHSH form~\cite{Clauser_1969}. For systems involving a higher number of particles, the construction of Bell inequalities becomes more intricate, with Mermin inequality being the most prominent example~\cite{MERMIN, Ardehali_1992, Belinskii1993}. 
More general constructions applicable to graph states have also been proposed in~\cite{Guhne_2005, Scarani_2005, Toth_2006, Guhne_2008, Baccari_2020}. Interestingly, tomographic measurement settings form a subset of the Clifford group, and therefore contain no non-stabilizer characteristics or \emph{quantum magic}—the very resource responsible for quantum computational advantage~\cite{Gottesman_1997, Gottesman_1998, Aaronson_2004}. 
Recently, it has been demonstrated that the violation of Bell inequalities is intrinsically connected with non-stabilizerness~\cite{Howard_2012, Howard_2014, Howard_2015}, and can serve as a witness of this resource~\cite{Macedo_2025, Cusumano_2025}. 
Consequently, our XYZ Bell inequalities derived for non-stabilizer states are promising candidates for certification of quantum magic and offer a deeper insight into the interplay between Bell nonlocality and quantum magic.

In this work, we introduce a constructive framework to derive several XYZ Bell inequalities and show its benefit for some of the most interesting $N$-qubit states ($N<6$). 
We analyze their performance, also in comparison with randomly chosen states, with the best known optimized inequalities by examining their noise robustness through the corresponding critical visibilities. 
Our results show that, on average, the constructed inequalities perform only marginally worse than the best possible ones. 
The presented framework enables any tomographic dataset--including previously collected experimental data--to be reinterpreted in terms of fundamental nonlocality tests at no additional experimental cost.
{\color{black} Here, we emphasize that conclusions about violations depend on the specific experimental details. Only when all assumptions of the Bell theorem are satisfied can one infer the non-existence of local hidden variable theories without loopholes.}
Finally, we explore the capability of the proposed inequalities to witness quantum magic by computing the stabilizer state maxima for those inequalities that are not maximally violated by pure graph states. We identify instances where such certification is possible and provide distinct results for both fixed Pauli measurement and general measurement setting scenarios, including experimental validation. 
Overall, our approach provides a method for the simultaneous certification of multiple quantum resources without requiring any additional experimental effort.

\section{Preliminaries}

Using the operator basis formed by the tensor products of Pauli matrices, any $N$-qubit quantum state can be represented as
\begin{equation}
    \rho=\frac{1}{2^N}\sum_{i_1,\ldots,i_N=0}^3 T_{i_1, \ldots i_N} \sigma_{i_1} \otimes \cdots \otimes \sigma_{i_N}, 
\end{equation}
where the $T_{i_1, \ldots i_N}$ form the so-called correlation tensor. It is of wide use in the Bell non-locality and entanglement detection and analysis~\cite{Horodecki_1995, Zukowski_2002, Nagata_2004, Hassan_2008, Badziag_2008, Tran_2015, Ketterer_2019, Wyderka_2020, Imai_2021, Ketterer_2022}. In this work, we are interested in Bell inequalities that are violated with Pauli measurements. Throughout this paper, we will refer to them as XYZ Bell inequalities. A general $N$-party $m$-setting Bell inequality is given as
\begin{eqnarray}
\label{eq:general_ineq}
    B(\vec{c}\,)=\sum_{i_1,\ldots,i_N=0}^{m+1} c_{i_1,\ldots,i_N} \, O_{1}^{(i_1)} \cdots O_{N}^{(i_N)} \leq L,
\end{eqnarray}
where $O_{k}^{(i_k)}$ is the $i$th observable for the $k$th party, with $O_{k}^{(0)}$ assumed to be an identity. The vector $\vec{c}\,$  is a vector of Bell coefficients, and $L$ denotes the local realistic bound. Clearly, in the considered scenario $m =2,3$, and we are interested in such inequalities and states $\rho$ for which the quantum expectation value $Q=\langle B(\vec{c}\,)\rangle_{\rho}$  yields a violation, i.e.,
\begin{equation}\label{eq:quantum_value}
    Q  = \sum_{i_1,\ldots,i_N \in \lbrace m \rbrace } c_{i_1,\ldots,i_N} \, T_{i_1, \ldots i_N} > L,
\end{equation}
where $\lbrace m \rbrace$ stands for the subset of Pauli matrices indices. For example, $\lbrace m \rbrace= \lbrace 1,2 \rbrace$ corresponds to the $xy$ plane settings and $\lbrace m \rbrace= \lbrace 1,2,3 \rbrace$ to the entire Pauli group with an identity matrix omitted in the notation. 

A useful quantity providing insight into the strength of nonlocality exhibited by a given state is the critical visibility $v_{\mathrm{crit}}$. It captures its robustness to white noise admixture and allows one to identify inequalities whose violation is experimentally feasible. For the correlation inequalities studied in this paper, it can be evaluated as $v_{\mathrm{crit}}=L/Q$.

A natural framework for finding the desired inequalities starts with employing linear programming techniques. Let $DS$ stand for the set of all deterministic strategies in the $N$-party $m$-setting scenario. For an $N$-qubit quantum state $\rho$ and the Pauli measurements subset $\lbrace m \rbrace$ we define the following linear programming problem
\begin{eqnarray}
    &&\max_{\vec{c}} Q\\
    \quad s.t.\, &&\max_{DS} \, \langle B(\vec{c}) \rangle \leq 1 \nonumber
\end{eqnarray}
If the maximization succeeds, it yields the set of coefficients $c_{i_1,\ldots,i_N}$ defining the XYZ Bell inequality with the local bound $L=1$. In the following section, we present the obtained inequalities--constructed using the method from~\cite{Cieslinski_2024poly}--for several $N$-qubit states that are both of theoretical interest and are commonly discussed in the literature. 
Note that the resulting inequalities are inherently basis dependent by construction. All studied states are listed explicitly in Appendix~\ref{app:states}. Although local transformations of the states may improve the performance of the obtained inequalities, our framework remains completely general and can be applied to any given state.

\section{XYZ Bell inequalities}
\subsection{Three qubits}

For two qubits, the XYZ Bell inequalities are violated by the standard Bell states expressed in the computational basis only after a suitable basis transformation. Moreover, all of the resulting inequalities are of the CHSH type. 
Therefore, the first interesting case we consider is that of three qubits. 
In this scenario, there exist two distinct classes of pure states that are inequivalent under stochastic local operations and classical communication (SLOCC): the GHZ state and the W state~\cite{Dur_2000}.

For both classes, the optimal XYZ inequality corresponds to the Mermin-type, also known as the MABK, inequality~\cite{MERMIN, Ardehali_1992, Belinskii1993}. Since we restrict our analysis to Pauli measurements, the inequalities will be expressed using 
\(\sigma_1, \sigma_2, \sigma_3 = X, Y, Z\), with subscripts explicitly denoting the corresponding parties. It is important to emphasize that the results presented below \emph{are not witnesses of Bell-type correlations}; rather, they constitute proper Bell inequalities where all of the X,Y,Z measurements can potentially be exchanged with arbitrary settings. For the GHZ state given as $(|000\rangle + |111\rangle)/\sqrt{2}$, the inequality takes the following form
\begin{equation}
X_1 X_2 X_3 - Y_1 Y_2 X_3 - Y_1 X_2 Y_3 - X_1 Y_2 Y_3 \leq 2.
\end{equation}
This inequality is maximally violated by the considered state with the critical visibility of $v_{\rm crit}^{\rm XYZ}=\frac{1}{2}$.

For the W state $(|100\rangle + |010\rangle + |001\rangle)/\sqrt{3}$, the optimal violation is obtained for settings in the $y-z$ plane ($X_i \to -Z_i$), and the inequality reads
\begin{equation}
-Z_1 Z_2 Z_3 + Y_1 Y_2 Z_3 + Y_1 Z_2 Y_3 + Z_1 Y_2 Y_3 \leq 2.
\end{equation}
The critical visibility for the W state is given as $\frac{2}{3}$. The highest violation of $4$ for the above inequality occurs not for a W state, where it is limited to $3$, but for the GHZ state expressed in the Pauli X basis: $(|{\rm+++}\rangle - |{\rm---}\rangle)/\sqrt{2}$, where $|\pm \rangle = (|0\rangle \pm |1\rangle)/\sqrt{2}$. 

\subsection{Four qubits}
\label{sec:four_qubits}
For four qubits, the variety of entangled states and the corresponding XYZ Bell inequalities becomes significantly richer. 
Exemplary four-qubit states together with the corresponding Bell operators are summarized in Table~\ref{tab:4_qubits}. For comparison, it also includes the best-known critical visibilities and the tomographic critical visibilities. Finally, the last column presents the maximal eigenstates of the Bell operators and critical visibilities computed through the corresponding eigenvalues, which can be interpreted as the maximal noise threshold for the fixed measurement settings.
\begin{center}
\begin{table*}
\begin{tabular}{|c|c|c|c|c|c|c|c|} \hline \hline
No & State & $m$  & $v_{\mathrm{crit}}^{\mathrm{opt}}$ &  Bell operator $\mathcal{B}$ corresponding to the inequality $\langle B \rangle \leq L$  & $L$ & $v_{\mathrm{crit}}^{\mathrm{XYZ}}$ & $v_{\mathrm{crit}}^{\max}, \, |\varphi_{max}\rangle$ \\ \hline\hline
1&$|\mathrm{GHZ}_4\rangle$ &2 & $\frac{1}{2\sqrt{2}}$ & 
$\begin{array}{c}X_1 X_2 X_3 X_4 - Y_1 Y_2 X_3 X_4 - Y_1 X_2 Y_3 X_4 - X_1 Y_2 Y_3 X_4 \\- Y_1 X_2 X_3 Y_4 - X_1 Y_2 X_3 Y_4 - X_1 X_2 Y_3 Y_4 + Y_1 Y_2 Y_3 Y_4 \end{array}$ &
4 &
$\frac{1}{2}$ &
$\frac{1}{2}, \,|\mathrm{GHZ}_4\rangle$ \\ \hline  
& $|\mathrm{GHZ}_4\rangle$  &3 & $\frac{1}{2\sqrt{2}}$ & 
as above &
2 &
$\frac{1}{2}$ &
$\frac{1}{2}, \,|\mathrm{GHZ}_4\rangle$ \\ \hline \hline 
2& $|\mathrm{W}_4\rangle$ &2 & $0.529$&
$\begin{array}{c}
-3(Z_1 - Z_2 - Z_3 - Z_4) \\ -Z_1 Z_2 - Z_1 Z_3 - Z_2 Z_3 - Z_1 Z_4 
- Z_2 Z_4 - Z_3 Z_4 \\
+2( Z_1 X_2 X_3 +  X_1 Z_2 X_3 +  X_1 X_2 Z_3 +  Z_1 X_2 X_4 \\
+  X_1 Z_2 X_4 +   Z_1 X_3 X_4 +  Z_2 X_3 X_4 +  X_1 Z_3X_4\\
+  X_2 Z_3 X_4 +  X_1 X_2 Z_4 +  X_1 X_3 Z_4 +  X_2 X_3 Z_4) \\
-Z_1 Z_2 Z_3 - Z_1 Z_2 Z_4 - Z_1 Z_3 Z_4 - Z_2 Z_3 Z_4\\
+ 2(Z_1 Z_2 X_3 X_4 +  Z_1 X_2 Z_3 X_4 +  X_1 Z_2 Z_3 X_4 +\\
Z_1 X_2 X_3 Z_4 +   X_1 Z_2 X_3 Z_4 + X_1 X_2 Z_3 Z_4) \\
-2 X_1 X_2 X_3 X_4-3 Z_1 Z_2 Z_3 Z_4  \end{array}$
&
9 &
$\begin{array}{c}\frac{9}{17}\\\approx 0.529\end{array}$ &
$\begin{array}{c} \frac{9}{8 \sqrt{2}+9} \approx 0.443\\ a_+|\mathrm{W}\rangle+a_-|\overline{\mathrm{W}}\rangle \\
a_{\pm} = \mp \frac{\sqrt{2 \pm \sqrt{2}}}{4} \end{array}
$ \\ \hline 
3& $|\mathrm{W}_4\rangle$ & 3 & $0.479$ &
$\begin{array}{c}
- 3 (Z_1 + Z_2 + Z_3 + Z_4) \\
- Z_1 Z_2 - Z_1 Z_3 - Z_2 Z_3 - Z_1 Z_4 - Z_2 Z_4 - Z_3 Z_4 \\
+ X_2 X_3 Z_1 
+ X_2 X_4 Z_1 
+ X_3 X_4 Z_1 
+ X_1 X_3 Z_2 \\
+ X_1 X_4 Z_2 
+ X_3 X_4 Z_2 
+ X_1 X_2 Z_3 
+ X_1 X_4 Z_3 \\
+ X_2 X_4 Z_3 
+ X_1 X_2 Z_4 
+ X_1 X_3 Z_4 
+ X_2 X_3 Z_4 \\
+ Y_2 Y_3 Z_1
+ Y_2 Y_4 Z_1 
+ Y_3 Y_4 Z_1 
+ Y_1 Y_3 Z_2 \\
+ Y_1 Y_4 Z_2 
+ Y_3 Y_4 Z_2
+ Y_1 Y_2 Z_3 
+ Y_1 Y_4 Z_3 \\
+ Y_2 Y_4 Z_3 
+ Y_1 Y_2 Z_4 
+ Y_1 Y_3 Z_4 
+ Y_2 Y_3 Z_4 \\
- Z_1 Z_3 Z_4 
- Z_2 Z_3 Z_4 
- Z_1 Z_2 Z_3 
- Z_1 Z_2 Z_4 \\
+ X_3 X_4 Z_1 Z_2 
+ X_2 X_4 Z_1 Z_3 
+ X_1 X_4 Z_2 Z_3 \\
+ X_2 X_3 Z_1 Z_4 
+ X_1 X_3 Z_2 Z_4 
+ X_1 X_2 Z_3 Z_4 \\
+ Y_3 Y_4 Z_1 Z_2 
+ Y_2 Y_4 Z_1 Z_3 
+ Y_1 Y_4 Z_2 Z_3 \\
+ Y_2 Y_3 Z_1 Z_4 
+ Y_1 Y_3 Z_2 Z_4 
+ Y_1 Y_2 Z_3 Z_4 \\
- X_1 X_2 X_3 X_4 
- Y_1 Y_2 Y_3 Y_4 
- 3 Z_1 Z_2 Z_3 Z_4 \end{array}$ &
9 &
$\begin{array}{c}\frac{9}{17}\\\approx 0.529\end{array}$&
$\begin{array}{c}\frac{9}{17}\approx 0.529\\|\mathrm{W}_4\rangle\end{array}$ \\ \hline\hline 
4& $|\mathrm{D}_4^2\rangle$ &2 & 0.471 &
$\begin{array}{c}
- Z_1 Z_2 X_3 X_4  -  Z_1 X_2 Z_3 X_4   -  X_1  Z_2 Z_3 X_4 \\ -   Z_1 X_2 X_3 Z_4 - X_1  Z_2 X_3 Z_4 -  X_1 X_2 Z_3 Z_4 +  2 Z_1 Z_2 Z_3 Z_4  \end{array}$
&
4 &
$\frac{2}{3}$ &
$\begin{array}{c} \frac{1}{2} \\ \frac{1}{2} |\mathrm{GHZ}_4\rangle -\frac{\sqrt{3}}{2} |\mathrm{D}_4^2\rangle\end{array}$ \\ \hline 
5& $|\mathrm{D}_4^2\rangle$ & 3 & $0.440$ &
$\begin{array}{c}
-X_3 X_4 Z_1 Z_2  - X_2 X_4 Z_1 Z_3  - X_1 X_4 Z_2 Z_3 \\- X_2 X_3 Z_1 Z_4 - X_1 X_3 Z_2 Z_4- X_1 X_2 Z_3 Z_4\\ 
- Y_2 Y_4 Z_1 Z_3 - Y_3 Y_4 Z_1 Z_2 - Y_1 Y_4 Z_2 Z_3  \\- Y_2 Y_3 Z_1 Z_4  - 
 Y_1 Y_3 Z_2 Z_4  - Y_1 Y_2 Z_3 Z_4 \\
 + 4 Z_1 Z_2 Z_3 Z_4
  \end{array}$
&
8 &
$\frac{2}{3}$&
$|\mathrm{D}_4^2\rangle$ \\ \hline \hline 
& $|\mathrm{L}_4\rangle$ & 2 & 0.446 &
$-$
&
$-$ &
$-$ &
$-$ \\ \hline
6& $|\mathrm{L}_4\rangle$ & 3 & 0.411 &
$\begin{array}{c}
 X_1 X_2 Z_3 + Z_1 X_3 X_4 + Z_2 X_3 X_4+  X_1 X_2 Z_4 \\
 - Y_1 Y_2 Z_3- Y_1 Y_2 Z_4- Z_1 Y_3 Y_4 - Z_2 Y_3 Y_4 \\
 + 2 (Y_1 X_2 Y_3 X_4 +  X_1 Y_2 Y_3 X_4 +  Y_1 X_2 X_3 Y_4 +  X_1 Y_2 X_3 Y_4) \end{array}$
&
8 &
$\frac{1}{2}$ &
$|\mathrm{L}_4\rangle$ \\ \hline  \hline
\end{tabular}
\caption{\label{tab:4_qubits}XYZ Bell inequalities for the selected four-qubit states. The second column lists the considered states, along with the number of settings $m$ and the best-known critical visibilities in the corresponding Bell scenarios (shown in the fourth column). Subsequently, the Bell operators constructed with fixed Pauli measurements and their respective local bounds $L$ are presented. Critical visibilities obtained for the tomographic settings are denoted by $v_{\mathrm{crit}}^{\mathrm{XYZ}}$. The final column displays the eigenstates $|\varphi_{\max} \rangle$ of the listed Bell operators $\mathcal{B}=\sum_i \lambda_i |\varphi_i\rangle \langle \varphi_i|$ corresponding to the maximal eigenvalue $\lambda_{\max}$, together with their critical visibilities $v_{\mathrm{crit}}^{\max}$ computed as $L/\lambda_{\max}$. A detailed discussion of the obtained inequalities is provided in the main text.}
\end{table*}
\end{center}

As expected, for the GHZ state and the $x–y$ plane settings, the optimal inequality is again of the Mermin type. 
Adding settings does not modify the form of the inequality. 
The tomographic critical visibility, $v^{\rm XYZ}_{crit}$ is suboptimal and equals $\frac{1}{2}$, while the optimal visibility of $\frac{1}{2\sqrt{2}}$ can be achieved within the MABK family of inequalities~\cite{Ardehali_1992, Belinskii1993}. The situation changes for the W state. 
In this case, the obtained tomographic inequality is more intricate, containing correlations of all orders, i.e., measurement terms involving one up to four parties, which consequently makes the tomographic violation as strong as the optimal one. However, the maximal violation of this inequality for fixed Pauli measurements is not achieved by the W state itself, but rather by a superposition of the W and $\overline{\mathrm{W}}$ states, where the bar denotes a state transformed by $X^{\otimes N}.$ 
For three settings per party, the critical visibility for Pauli measurements remains unchanged, while the optimal value decreases. 
Interestingly, going beyond in-plane settings causes the W state to yield the highest XYZ violation—that is, it becomes the eigenvector of the Bell operator corresponding to its maximal eigenvalue.  
For the Dicke state $|\mathrm{D}^2_4\rangle,$ the XYZ inequalities are less robust to noise than the optimal ones. 
Here as well, extending the measurement settings beyond the plane does not improve $v_{\mathrm{crit}}^{\rm XYZ},$ but it does make the Dicke state the maximally violating one. 

A different behaviour arises for the linear cluster state $|\mathrm{L}_4\rangle.$ A two-setting Pauli-measurement inequality does not exist. 
Only a suitable local unitary transformation would yield a violation which, however, is in conflict with our no-cost resource certification approach. For three settings, an inequality with a critical visibility of $\frac{1}{2}$, maximally violated by the $|\mathrm{L}_4\rangle$ state, was found. The optimal critical visibility is slightly lower, while the tomographic one coincides with the stabiliser construction reported in~\cite{Guhne_2005}. The optimal critical visibilities were computed using the linear programming technique described in~\cite{Gruca_2010}. 

\begin{figure}
\centering
\includegraphics[width=0.42\textwidth]{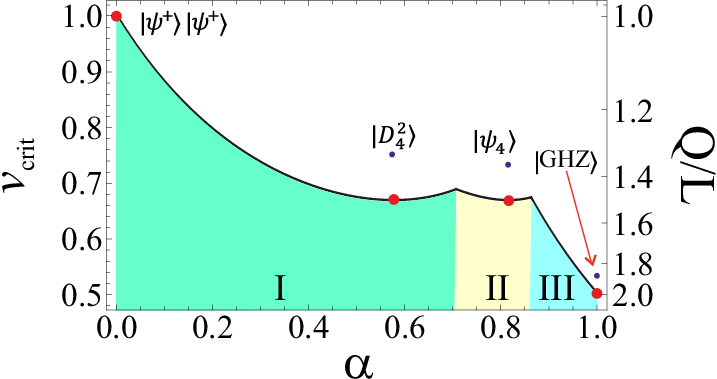}
\caption{Optimal critical visibility of the $|\psi(\alpha)\rangle$ states as a function of $\alpha$ (solid line and red points) for the XYZ Bell inequalities presented in this work together with the experimental violation factors (blue points). The coloured regions marked with Roman numbers correspond to different optimal inequalities which vary depending on the state's parameter. For region I, which encompasses the product of two-qubit Bell states and Dicke state $|\mathrm{D}^2_4\rangle$, the optimal inequality is presented in Table ~\ref{tab:4_qubits} under \# 5. By further increasing $\alpha$, we switch to region II containing the four-qubit state $|\psi_4\rangle$ with the corresponding optimal inequality (\ref{ineqpsi4}). The last region labelled by III is governed by the Mermin inequality (see \# 1 in Table~\ref{tab:4_qubits}). For analytic expressions, see the main text. The right axis on the above plot denotes the violation factors $Q/L$. Experimental data for the chosen states gathered in \cite{PhysRevLett.117.210504, PhysRevLett.107.080504,Schmid_2010} are represented by the blue points, with the
diameter representing the estimated error, and lead to a clear violation.}
\label{fig:dino}
\end{figure}

Now, we will focus on an especially interesting example of a whole family $\ket{\psi(\alpha)}$ of four-qubit quantum states~\cite{Wieczorek_2008, Schmid_2008phd} defined as
\begin{eqnarray}
    |\psi(\alpha)\rangle&=& \frac{\alpha}{\sqrt{2}}( |0011\rangle + |1100\rangle)\\&+&  \frac{\sqrt{1-\alpha^2}}{2} (|0101\rangle+ |0110\rangle+| 1001\rangle+| 1010\rangle) \nonumber 
    \label{eq:psi_alpha}
\end{eqnarray}
which, for different values of the parameter $\alpha$, includes states such as $|\mathrm{GHZ}_4\rangle$, $|\mathrm{D}_4^2\rangle$, $|\psi_4\rangle$, and $|\psi^+\rangle |\psi^+\rangle$. Here, $|\psi_4\rangle$ denotes the four-qubit  state $\psi(\alpha = \sqrt{2/3})$ directly obtained from parametric down conversion~\cite{Weinfurter_2001, Eibl_2003} (see Appendix~\ref{app:states}), and $|\psi^+\rangle = (|01\rangle + |10\rangle)/\sqrt{2}$ is one of the two-qubit Bell states. 
Remarkably, for the whole range of $\alpha$ we find only three XYZ Bell inequalities where then only critical visibility directly depends on $\alpha$ (see Fig.~\ref{fig:dino}). The inequality characteristic to the state $|\mathrm{D}_4^2\rangle$ (see $\#5$ in Table~\ref{tab:4_qubits}) extends to all values of the parameter $0 \leq \alpha \leq \frac{1}{\sqrt{2}}$, and is violated with a critical visibility
\begin{equation}
v^{\mathrm{XYZ}}_{\mathrm{crit}}= \frac{1}{-\frac{\alpha ^2}{2}+\sqrt{2-2 \alpha ^2} \alpha +1}.
\end{equation}
For all states with $\frac{1}{\sqrt{2}} < \alpha \leq \sqrt{\frac{7+4\sqrt{2}}{17}}$, the inequality takes the form
\begin{eqnarray}
&2 Z_1 Z_2 Z_3 Z_4& \nonumber \\
&- Y_1 Y_2 X_3 X_4 + Y_1 X_2 Y_3 X_4 + X_1 Y_2 Y_3 X_4& \nonumber \\
&+ Y_1 X_2 X_3 Y_4 + X_1 Y_2 X_3 Y_4 - X_1 X_2 Y_3 Y_4 &\nonumber \\
&- Z_1 Z_2 X_3 X_4 - Z_1 X_2 Z_3 X_4 - X_1 Z_2 Z_3 X_4 & \label{ineqpsi4}\\
&- Z_1 X_2 X_3 Z_4 - X_1 Z_2 X_3 Z_4 - X_1 X_2 Z_3 Z_4& \nonumber \\
&- Z_1 Z_2 Y_3 Y_4 - Z_1 Y_2 Z_3 Y_4 - Y_1 Z_2 Z_3 Y_4& \nonumber \\
&- Z_1 Y_2 Y_3 Z_4 - Y_1 Z_2 Y_3 Z_4 - Y_1 Y_2 Z_3 Z_4 \leq 8&\nonumber
\end{eqnarray}
and is violated with critical visibility:
\begin{equation}
v^{\mathrm{XYZ}}_{\mathrm{crit}} = \frac{2}{\alpha ^2+2 \sqrt{2-2 \alpha ^2} \alpha +1}.
\end{equation}
For the final range of the parameter $\sqrt{\frac{7+4\sqrt{2}}{17}} < \alpha \leq 1$, which includes the $|\mathrm{GHZ}\rangle$ state, the appropriate inequality is the Mermin inequality ($\#1$ in Table~\ref{tab:4_qubits}), which is violated with critical visibility of $v^{\mathrm{XYZ}}_{\mathrm{crit}} = \frac{1}{2 \alpha ^2}$. Violation factors extracted from the experimental data obtained in \cite{PhysRevLett.117.210504, PhysRevLett.107.080504,Schmid_2010} (see Table~\ref{tab:exp_violation}) for the chosen four-qubit states are also shown in the Fig.~\ref{fig:dino} (blue points).

\subsection{Five qubits}

In the case of five qubits, we have considered both the previously studied states and two additional examples: a graph state corresponding to a closed (ring) graph, denoted by $|\mathrm{R}\rangle$, and an absolutely maximally entangled (AME) state (see Appendix~\ref{app:states} for the explicit expressions). 
It is worth emphasising that our approach can be applied to \emph{any} quantum state. The cases presented here serve as illustrative examples of its use. 

The obtained XYZ Bell inequalities are too long to be presented explicitly in the main text. For completeness, we provide them in a separate supplementary file in a convenient, ready-to-use format available in Supplemental Materials. The corresponding critical visibilities are provided in Table~\ref{tab:5_qubits_v}). The highest noise robustness is observed for the $|\mathrm{D}^2_5\rangle$ state. For all of the states, except GHZ, the two-setting Bell inequalities are based on the $X$ and $Z$ measurements.

\begin{table}
     \centering
   \begin{tabular}{|c|c|c|}
\hline\hline
\multirow[t]{2}{*}{State} & \multicolumn{2}{c|}{$v^{\mathrm{XYZ}}_{\mathrm{crit}}$} \\
\cline{2-3}
                          & $m=2$ & $m=3$ \\
\hline
         $|{\rm GHZ}_5\rangle $  & $0.250$ & $0.250$ \\
         $|W_5\rangle $  & $0.250$ & $0.143$ \\
         $|D^2_5\rangle $  & $0.111$ & $0.111$\\
         $|L_5\rangle $  & $0.333$ & $0.143$ \\
         $|R_5\rangle $  & $0.600$ & $0.246$ \\
         $|{\rm AME}(5,2)\rangle$  & $0.333$ & $0.246$ \\
         \hline \hline
    \end{tabular}
    \caption{\label{tab:5_qubits_v} Critical visibilities for the studied five-qubit states and the corresponding XYZ Bell inequalities, compared with their optimal values. The first column lists the states under consideration. 
The next two columns provide the corresponding tomographic critical visibilities $v^{\mathrm{XYZ}}_{\mathrm{crit}}$ obtained using fixed Pauli measurements.}
    \end{table}

\subsection{Comparison with the optimal inequalities and settings for $N=4$} 

XYZ Bell inequalities are, in general, not expected to yield the lowest best known critical visibilities $v_{\mathrm{crit}}$ for a given state and scenario. However, when compared with the best known optimized inequalities, their performance proves to be surprisingly good.
This already becomes apparent comparing the critical visibilities for the studied four-qubit states presented in Table~\ref{tab:4_qubits}. To make more general statements about their robustness against noise, we analyse $v^{\mathrm{XYZ}}_{\mathrm{crit}}$ in the tomographic setting with $m=3$ and compare it with the minimal $v^{\mathrm{opt}}_{\mathrm{crit}}$ for randomly chosen states.

\begin{figure}[h!]
\label{fig:compare_v}
\includegraphics[width=0.45\textwidth]{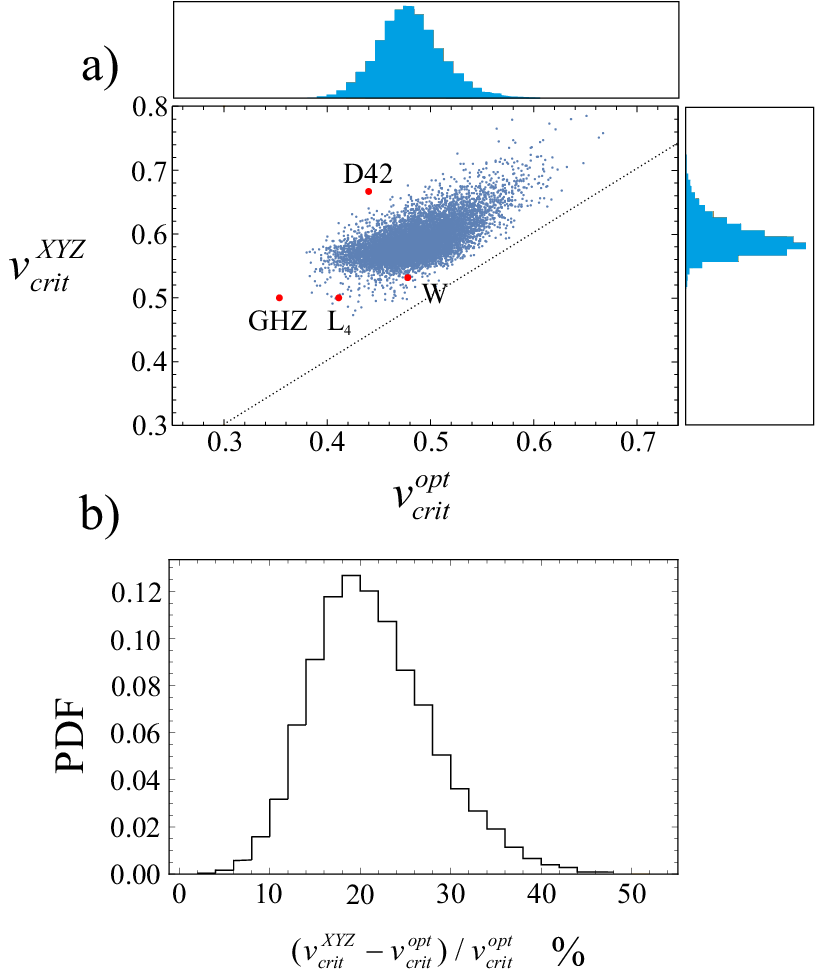}
\caption{Generic critical visibilities in XYZ and optimal Bell inequalities. Panel $a)$ shows the XYZ inequality's critical visibilities, and the optimal critical visibilities for $10^5$ pure Haar random states for $m=3$. Histograms of the marginal distributions are shown in blue on the corresponding sides of the plot. Results for the states previously studied in section~\ref{sec:four_qubits} are denoted by red points. Panel $b)$ shows an estimated probability density function of a relative difference between the critical visibilities from the upper panel. It shows that for a generic state, the $v^{\mathrm{XYZ}}_{\mathrm{crit}}$ for Pauli measurements is on average $\approx 20\%$ worse than the $v^{opt}_{\mathrm{crit}}$. The loss of visibility with such a constrained class of measurement settings is surprisingly low. 
}
\end{figure}

Our analysis combines the linear programming method introduced in~\cite{Gruca_2010} with techniques developed in this work. 
Specifically, we sampled $10^5$ pure random four-qubit states according to the Haar measure and, for each of them, determined the corresponding XYZ Bell inequality. 
Subsequently, we numerically optimized all of the states to identify the inequalities and settings that lead to the lowest achievable critical visibility. 
The results are shown in Fig.~\ref{fig:compare_v}. 
We find that, on average, the critical visibility obtained with Pauli measurements is approximately $20\%$ higher (worse) than the fully optimised value. The smallest observed relative difference in the simulation was $2.574\%$, while the largest reached $50.263\%$. Considering the simplicity and universality of tomographic measurement settings, their performance can be regarded as remarkably good.

\section{Stabilizer maxima and quantum magic witnessing}

Quantum magic, also referred to as non-stabilizerness, is a resource responsible for the advantage of quantum computations over classical ones. It stems from the Gottesman–Knill theorem, which states that any circuit that starts from a $|0\rangle^{\otimes N}$ state and consists only of Clifford operations can be efficiently simulated on a classical computer~\cite{Gottesman_1997, Gottesman_1998, Aaronson_2004}. Therefore, additional resources in the form of non-stabilizer states and non-Clifford gates are needed for the quantum improvement~\cite{Bravyi_2005, Bravyi_2016, Howard_2017}. The resource theory of quantum magic has found many applications~\cite{Liu_2022, Salvatore_2022, White_2024, Tirrito_2024, Turkeshi_2025, Brokemeier_2025, Hoshino_2025, Odavi_2025, Hou_2025, Catalano_2025, Junior_2025}, and its detection and understanding have become a fundamental problem on the path towards unraveling the computational potential of quantum systems.

Recently, the possibility of detecting quantum magic with Bell inequalities was proposed in \cite{Macedo_2025, Cusumano_2025}. Our scenario, involving only Pauli measurements, fits naturally into this picture as it does not introduce any non-Clifford resources to it. Witnessing magic with Bell inequalities can be studied in several ways. First, one can look for inequalities which, maximized over all stabilizer states and general measurement settings, yield non-trivial bounds that can be violated in quantum theory~\cite{Macedo_2025}. In the qubit case, such a task was performed there for at most three parties. Complementarily, one could take resource-free states and measurements and look at how Clifford and non-Clifford operations change their behavior~\cite{Cusumano_2025}. We note that both of these approaches provide new insight into the interplay of Bell nonlocality and non-stabilizerness.  Here, we will start with the resource-free device-dependent measurements approach and later consider optimization over all of the possible measurements and pure graph states. As a result, we will provide new instances of multipartite Bell inequalities capable of quantum magic witnessing, also supported by experimental data. {\color{black} It is worth noting that the set of non-stabilizer states, i.e., states lying outside the stabilizer set is not a proper subset of the set of states that violate Bell inequalities. However, there exists an overlap between these sets, which we exploit to draw two independent conclusions that hold simultaneously: the underlying state violates a Bell inequality and is not a stabilizer state.} 

\subsection{Stabilizer maxima for tomographic settings}

Let us consider a Bell operator $B(\vec{c})$ defined in (\ref{eq:general_ineq}). Any possible extreme expectations on the set of stabilizer states with fixed Pauli measurements have the following form 
\begin{equation}
    \langle B \rangle_{\mathrm{STAB}}\leq \sum_{i_1, \cdots, i_N=0,1,2}c_{i_1, \cdots, i_N} S_{i_1, \cdots, i_N},
\end{equation}
where $S_{i_1, \cdots, i_N}\in \lbrace -1,0,1 \rbrace$ and correspond to perfect correlations or no correlations at all, i.e. $T_{i_1, \ldots i_N}=\pm1,0$ in (\ref{eq:quantum_value}). To find the stabilizer maxima, we group the operators $O^{i_1}_1 \cdots O^{i_N}_N$ (Pauli strings) into the sets of mutually commuting observables by generating a commutativity graph $g$ and finding its maximal cliques $\omega(g)$ (i.e. we discard the cliques which are elements of a larger clique). The reason for that is that only these operators can simultaneously reach $\pm 1$ expectation values. For each of them, we assign all possible sign combinations from $\lbrace 1,-1 \rbrace^{\omega(g)}$ and check whether there exists a state stabilized by these operators. 
Having found all the stabilizer states, we compute $\langle B \rangle_{\mathrm{STAB}}$ and check its maximal value. An alternative approach can be performed by taking into account all pure stabilizer states, i.e. graph states $|G\rangle$, and optimizing the Bell operator over the local Clifford operations. 

We performed the above analysis for the four-qubit $|\mathrm{W}_4\rangle$ and $|\mathrm{D}^2_4\rangle$ states and the three-qubit Hoggar state~\cite{Howard_2017, Hoggar_1998}
\begin{eqnarray}
    |\mathrm{Hoggar}\rangle &=&\frac{1}{\sqrt{6}}[(1+i)|000\rangle - |010\rangle \nonumber \\
    &+& |011\rangle -i|100\rangle + |101\rangle] 
    \label{eq:hoggar}
\end{eqnarray}
It is worth noting that the latter is a maximally magical three-qubit state, which, contrary to the two-qubit case, is entangled. The corresponding three-setting inequality for the Hoggar state is
\begin{eqnarray}
\label{eq:hoggar_ineq}
&-&14 X_1 + 61 Y_1 - 8 Z_1 - 17 X_2 -20 Y_2  \\
&+& 2 Z_2+ 2 X_3- 67 Y_3+ 8 Z_3 \nonumber \\
&+& 50 X_1 X_2 + 21 Y_1 X_2 - 12 Z_1 X_2 - 9 X_1 Y_2  + Z_1 Y_2 \nonumber \\
&+& 34 Y_1 Y_2 - 27 X_1 Z_2 - 30 Y_1 Z_2 + 9 X_1 X_3- 37 Y_1 X_3 \nonumber \\
&-& 55 Z_1 Z_2 + 2 Z_1 X_3 + 8 Y_2 X_3+ 52 Z_2 X_3+ 3 X_1 Y_3 \nonumber \\
&+& 24 X_2 X_3 - 22 Z_1 Y_3 - 6 X_2 Y_3- 34 Y_2 Y_3 - 24 Y_1 Y_3\nonumber \\
&+& 33 Z_2 Y_3- 20 X_1 Z_3 + 6 Y_1 Z_3 + 18 Z_1 Z_3+ 6 Y_2 Z_3 \nonumber \\
&-& 47 X_2 Z_3+ 33 Z_2 Z_3 \nonumber \\
&-& 12 X_1 X_2 X_3 - 41 Y_1 X_2 X_3 + 53 Z_1 X_2 X_3  + 47 X_1 Y_2 X_3 \nonumber \\
&-& 23 Y_1 Y_2 X_3 - 78 Z_1 Y_2 X_3  + 26 X_1 Z_2 X_3 - 55 Y_1 Z_2 X_3 \nonumber \\
&+& 23 Z_1 Z_2 X_3 +  15 X_1 X_2 Y_3 + 47 Y_1 X_2 Y_3 - 56 Z_1 X_2 Y_3 \nonumber \\
&-& 44 X_1 Y_2 Y_3 + 55 Y_1 Y_2 Y_3 - 23 Z_1 Y_2 Y_3  - 56 X_1 Z_2 Y_3 \nonumber \\
&-& 78 Y_1 Z_2 Y_3 - 55 Z_1 Z_2 Y_3 - 117 X_1 X_2 Z_3 + 21 Y_1 X_2 Z_3 \nonumber \\
&-& 15 Z_1 X_2 Z_3  - 12 X_1 Y_2 Z_3 - 44 Y_1 Y_2 Z_3 + 56 Z_1 Y_2 Z_3 \nonumber \\
&-& 3 X_1 Z_2 Z_3 - 53 Y_1 Z_2 Z_3 + 23 Z_1 Z_2 Z_3 \leq 429 \nonumber
\end{eqnarray}
with $v_{crit}^{\mathrm{XYZ}}=33/41$. The obtained stabilizer bounds together with quantum values for all three states and Pauli measurements only are presented in Table~\ref{tab:stab_max}. Clearly, the stabilizer maxima in all but one case are smaller than the quantum mechanically achievable values, and \emph{given the Pauli measurements, one can claim quantum magic witnessing if the violation is observed}. With no assumptions on the measurement settings, such a claim is still possible; however, it is not achievable by the studied states, as will be shown in the next subsection.

\begin{table}[]
    \centering
    \begin{tabular}{|c|c|c|c|c|}
    \hline
    \hline
         Inequality & $m$ & State  & XYZ state  & XYZ STAB \\
         \hline
         \text{\#2}  &2 & $|\mathrm{W}_4\rangle$ & $17$ & $17$ \\
         \text{\#3}  &3& $|\mathrm{W}_4\rangle$ & $17$ & $13$   \\
         \text{\#5}  &3& $|\mathrm{D}_4^2\rangle$ & $12$ & $10$  \\
         (\ref{eq:hoggar_ineq}) &3 &  $|\mathrm{Hoggar}\rangle$ & $533 $ & $392$     \\
         \hline
         \hline
    \end{tabular}
    \caption{ \label{tab:stab_max}Stabilizer maxima analysis for the generated Bell inequalities and Pauli measurements only. The first and the second column refers to the inequality under consideration and the corresponding number of settings respectively. Its maximal value with Pauli measurements reached by the generating state is given in the forth column. The corresponding stabilizer maximum under the $X,Y,Z$ measurements is denoted as XYZ STAB.}
    \end{table}

\subsection{Stabilizer maxima for arbitrary settings}

For completeness, we have also maximized our inequalities over the set of pure stabilizer (graph) states and any possible settings. This approach corresponds exactly to the magic witnessing from~\cite{Macedo_2025}. However, here we go beyond three parties and perform our analysis for more states. Also, we perform the optimization over measurement settings and states via semidefinite programming (SDP) using the standard see-saw technique.  

In the case of two settings per party, we have considered the inequality generated for a four-qubit W state. Its maximal violation for arbitrary settings is given as $17$. Note that this value is achievable through the Pauli measurements. Analogously, by optimizing the same inequality over all four-qubit graph states and measurements, we obtained the stabilizer bound of $17$, which is again the W state maximum. Nevertheless, the quantum maximum, i.e. maximum quantum value over all four-qubit states and measurements, yields $8 \sqrt{2}+9\approx 20.3137$. Thus, inequality \#2 can serve as a magic witness. However, the violation of the stabilizer bound is not observed for the W state. 

For the $|\mathrm{W}_4\rangle$ state inequality and three settings per party, the optimized stabilizer and quantum maxima are again $17$ and $20.3137$, respectively. The maximal $|\mathrm{W}_4\rangle$ violation is again equal to $17$. For the $|\mathrm{D}^2_4 \rangle$ inequality, they are both given as $16$ and the generating state achieves only $12.056$. In the case of the Hoggar state generated inequality, the quantum maximum is $706.512$, with the stabilizer bound of $619.021$. The highest possible violation with the Hoggar state is the same as the stabilizer bound.  
All results presented in this section are summarized in Table \ref{tab:stab_max_Q}.

In conclusion, XYZ Bell inequalities obtained in this work can be used for magic witnessing in both fixed stabilizer measurements as well as arbitrary measurement scenarios; however, the inequalities are not violated by the generating states in the latter case. In the subsequent section, we will provide experimental data yielding a violation of a stabilizer bound in the XYZ setting.

\begin{table}[]
    \centering
    \begin{tabular}{|c|c|c|c|c|c|}
    \hline
    \hline
         Inequality & $m$ & State  & Q state  & STAB & Q \\
         \hline
         \text{\#2}  &2 & $|\mathrm{W}_4\rangle$ & $17$ & $17$ & $20.314$ \\
         \text{\#3}  &3& $|\mathrm{W}_4\rangle$ & $17$ & $17$  & $20.314$ \\
         \text{\#5}  &3& $|\mathrm{D}_4^2\rangle$ & $12.056$ & $16$ & $16$  \\
         (\ref{eq:hoggar_ineq}) &3 &  $|\mathrm{Hoggar}\rangle$ & $619.021 $ & $619.021$ & $706.512$     \\
         \hline
         \hline
    \end{tabular}
    \caption{\label{tab:stab_max_Q} Quantum and stabilizer  maxima analysis for the generated Bell inequalities. The relevant inequalities, corresponding number of settings and generating states are provided in the first three columns. The highest violation achievable by the generating state for arbitrary settings is denoted as Q state. Results obtained through the optimization over all measurements and all graph states (qubit states) are presented as STAB (Q). Whenever Q $>$ STAB, the inequality can serve as a magic witness independent of the measurement settings. All of the above values were computed using see-saw SDP optimization.}
    \end{table}

\section{Experimental data}

We present the usefulness of the XYZ Bell inequalities by reanalyzing data from four-qubit state tomography experiments performed previously ~\cite{PhysRevLett.117.210504, PhysRevLett.107.080504,Schmid_2010}. We use the correlations observed during the state tomography, not the finally obtained density matrices. From these values (Table~\ref{tab:exp_correlations}) we calculate the violation factors $Q/L$ for the respective XYZ Bell inequalities. All of them clearly prove the Bell nonlocality of the experimentally observed states (Table~\ref{tab:exp_violation}). {\color{black} Importantly, the interpretation of the observed value of the Bell expression strongly depends on the specific experimental realization. If all assumptions of the Bell theorem are satisfied, one can infer the strong conclusion regarding the violation of local realism. Otherwise, only the verification of Bell-like correlations is possible.}

{\color{black} In addition to the standard loopholes present in any Bell test, our work may be affected by the limitation arising from the fixed and predetermined choice of measurement settings. However, this can be readily overcome by introducing random switching between XYZ measurements in future state reconstruction experiments, as this does not increase the total number of measurements, albeit, admittedly, it may increase the total measurement time in some experimental implementations due to the frequent changes of measurement settings. Independent of the above considerations, the proposed Bell expressions can be used to draw conclusions about the presence of magic in the examined state.} The quantum value of $10.696$ in the case of the $|D^2_4 \rangle$ state and inequality \#5 was strong enough to violate the fixed Pauli measurements stabilizer maxima of 10, and thus signals the quantum magic resources witnessed experimentally through Bell {\color{black} inequality violation.}

\begin{table}
\begin{tabular}{|c | c | c |c |c|} \hline \hline
Correlation &$|\mathrm{D}_4^2\rangle$&$|\mathrm{GHZ}_4\rangle$&$|\mathrm{L}_4\rangle$&$|\psi_4\rangle$ \\ \hline
$\langle X_1X_2Z_3 \rangle$ &&&0.942& \\
$\langle X_1X_2Z_4 \rangle$ &&&0.944& \\
$\langle Z_1X_3X_4 \rangle$ &&&0.981& \\
$\langle Z_2X_3X_4 \rangle$ &&&0.981& \\
$\langle Y_1Y_2Z_3 \rangle$ &&&-0.950& \\
$\langle Y_1Y_2Z_4 \rangle$ &&&-0.947& \\
$\langle Z_1Y_3Y_4 \rangle$ &&&-0.980& \\
$\langle Z_2Y_3Y_4 \rangle$ &&&-0.980& \\
$\langle X_1X_2X_3X_4 \rangle$ &&0.939&&0.891 \\
$\langle X_1X_2Y_3Y_4 \rangle$ &&-0.932&&-0.317 \\
$\langle X_1X_2Z_3Z_4 \rangle$ &-0.595&&&-0.299 \\
$\langle X_1Y_2X_3Y_4 \rangle$ &&-0.928&0.925&0.602 \\
$\langle X_1Z_2X_3Z_4 \rangle$ &-0.620&&&-0.585 \\
$\langle X_1Y_2Y_3X_4 \rangle$ &&-0.925&0.944&0.562 \\
$\langle X_1Z_2Z_3X_4 \rangle$ &-0.570&&&-0.623 \\
$\langle Y_1X_2X_3Y_4 \rangle$ &&-0.928&0.935&0.572 \\
$\langle Z_1X_2X_3Z_4 \rangle$ &-0.602&&&-0.645 \\
$\langle Y_1X_2Y_3X_4 \rangle$ &&-0.921&0.930&0.626 \\
$\langle Z_1X_2Z_3X_4 \rangle$ &-0.562&&&-0.588 \\
$\langle Y_1Y_2X_3X_4 \rangle$ &&-0.937&&-0.278 \\
$\langle Z_1Z_2X_3X_4 \rangle$ &-0.597&&&-0.275 \\
$\langle Y_1Y_2Y_3Y_4 \rangle$ &&0.933&&0.902 \\
$\langle Y_1Y_2Z_3Z_4 \rangle$ &-0.595&&&-0.343 \\
$\langle Y_1Z_2Y_3Z_4 \rangle$ &-0.569&&&-0.574 \\
$\langle Y_1Z_2Z_3Y_4 \rangle$ &-0.598&&&-0.610 \\
$\langle Z_1Y_2Y_3Z_4 \rangle$ &-0.576&&&-0.621 \\
$\langle Z_1Y_2Z_3Y_4 \rangle$ &-0.563&&&-0.595 \\
$\langle Z_1Z_2Y_3Y_4 \rangle$ &-0.616&&&-0.291 \\
$\langle Z_1Z_2Z_3Z_4 \rangle$ &0.909&&&0.982 \\ \hline \hline
\end{tabular}
\caption{Explicit experimental values of the correlation measurements (Pauli strings) from~\cite{PhysRevLett.117.210504, PhysRevLett.107.080504,Schmid_2010} constituting the examined XYZ Bell inequalities. {\color{black} The average number of counts per correlation measurement for each of the above states was 1846, 3832, 4406, and 921, respectively.} \label{tab:exp_correlations}}
\end{table}

\begin{table}
\begin{tabular}{| c | c | c | c|} \hline \hline
State & Inequality & Violation factor & Experiment \\ \hline
$|\mathrm{GHZ}_4\rangle$ & $\#1$ & $1.861 \pm 0.011$& \cite{PhysRevLett.117.210504}\\
$|\mathrm{D}_4^2 \rangle$ &$\#4$& $1.341 \pm 0.018$ &\cite{PhysRevLett.107.080504}\\
 &$\#5$ & $1.337 \pm 0.015$& \cite{PhysRevLett.107.080504}\\
$|\mathrm{L}_4\rangle$ & $\#6$ & $1.897 \pm 0.009$& \cite{PhysRevLett.117.210504}\\
$|\psi_4\rangle$ &(\ref{ineqpsi4})& $1.357 \pm 0.018$& \cite{Schmid_2010}\\\hline \hline 
\end{tabular}
\caption{Experimental violation factors $Q/L$ for the chosen states and XYZ Bell inequalities studied in this work. Respective quantum values were evaluated using experimental data of Pauli observables measurements (see Tab.~\ref{tab:exp_correlations}) on the entangled photon states. \label{tab:exp_violation}}
\end{table}

\section{Conclusions}
In this work, we have constructed Bell inequalities that rely only on tomographic measurements, performed through the measurements in the Pauli basis, which can be used to certify violations of Bell inequalities at no additional experimental cost. We introduce a generic framework to generate the respective XYZ Bell inequalities. We have shown that such a constrained set of measurement settings does not perform much worse, in terms of robustness against experimental noise, than the optimized one in a generic scenario. We analyzed the obtained inequalities from the magic witnessing perspective and showed that they can be used for this task. 
Our XYZ Bell inequalities allowed reinterpreting the already
existent tomographic type data in terms of fundamental
nonlocality tests or certifying the nonlocality of the observed states for quantum information processing.
The examined experimental data supported a positive conclusion on the magic witnessing with fixed non-stabilizer measurement settings for the $|D_4^2\rangle$ state.

The complexity to obtain the relevant data, as well as to generate the inequalities, increases exponentially with the number of qubits. Yet, the method presented in this paper is of direct use for any qubit state up to 5 parties and can be employed for any experimental data where the Pauli measurements were performed. Given sufficient computational resources, the proposed technique can be extended to a higher number of parties and dimensions. XYZ Bell inequalities indeed can serve as a certificate of the genuine quantumness of the prepared state as they signal entanglement, Bell nonlocality and, in some cases, also the quantum magic.

\section*{Acknowledgements}
We thank Felix Huber and Gerard Angl{\'e}s Munn{\`e} for helpful discussions on the stabiliser and quantum maxima.  PC acknowledges the support of the Foundation for Polish Science (FNP) within the START programme.  WL is supported by the National Science Centre (NCN, Poland) within the OPUS project (Grant No. 2024/53/B/ST2/04103). We acknowledge funding from the German Federal Ministry of Research, Technology, and Space (Bundesministerium für Forschung, Technik und Raumfahrt, BMFTR) within project QuKuK (Contract No. 16KIS1621) and from the German Research Foundation (Deutsche Forschungsgemeinschaft, DFG) under Germany’s Excellence Strategy-EXC-2111-390814868. This work is partially carried out under IRA Programme, project no. FENG.02.01-IP.05-0006/23, financed by the FENG program 2021-2027, Priority FENG.02, Measure FENG.02.01., with the support of the FNP.

\appendix

\section{List of the studied states}
\label{app:states}
Here, we provide the explicit forms of the quantum states studied in this work. We follow the standard notation in which $|0\rangle$ and $|1\rangle$ denote the $+1$ and $-1$ eigenstates of the $Z$ operator, respectively. The $N$-qubit GHZ state is defined as
\begin{equation}
    |\mathrm{GHZ}_N\rangle =\frac{1}{\sqrt{2}}(|0\rangle ^{\otimes N}+|1\rangle ^{\otimes N})
\end{equation}
The $|\mathrm{W}_N\rangle$ state is given as
\begin{equation}
    |\mathrm{W}_N\rangle =\frac{1}{\sqrt{N}}(|10\cdots 0\rangle + |010\cdots 0\rangle +\cdots |0\cdots 01 \rangle)
\end{equation}
Its generalisation to $k$ excitations is given by the $|D^k_N\rangle$ state
\begin{eqnarray}
    |\mathrm{D}^k_N\rangle &=&{N \choose k}^{-1/2}(|\underbrace{11\cdots 1}_{k} 0\cdots 0\rangle + |10\underbrace{1\cdots1}_{k-1} 0\cdots 0\rangle \nonumber \\
    &+&\cdots |0\cdots 0\underbrace{1\cdots1}_{k} \rangle)
\end{eqnarray}
For $N=4,5$ the following graph states were investigated
\begin{eqnarray}
    |\mathrm{L}_4\rangle &=& \frac{1}{2}(|0000\rangle + |0011\rangle + 
    |1100\rangle - |1111\rangle) \\
 |\mathrm{R}_5\rangle &=& \frac{1}{4}(
 -|00001\rangle -|00010\rangle -|00100\rangle +|00111\rangle \nonumber \\
 &-&|01000\rangle -|01011\rangle -|01101\rangle +|01110\rangle \nonumber \\
 &-&|10000\rangle +|10011\rangle -|10101\rangle -|10110\rangle\nonumber \\
 &+&|11001\rangle -|11010\rangle +|11100\rangle +|11111\rangle), \\
|\mathrm{L}_5\rangle &=& \frac{1}{4}(
 |00000\rangle +|00010\rangle +|00101\rangle -|00111\rangle  \nonumber \\&+&|01000\rangle +|01010\rangle +|01101\rangle -|01111\rangle \nonumber \\
&+&|10001\rangle -|10011\rangle +|10100\rangle +|10110\rangle\nonumber \\
 &-&|11001\rangle +|11011\rangle -|11100\rangle -|11110\rangle).
\end{eqnarray}
Additionally, the four-qubit singlet state and an absolutely maximally entangled state of five qubits were taken into account
\begin{eqnarray}
    |\psi_4\rangle &=& \psi(\alpha=\sqrt{2/3})\\
 |\mathrm{AME}_5\rangle &=& \frac{1}{2\sqrt{2}}(
 -|00000\rangle -|00110\rangle -|01001\rangle +|01111\rangle \nonumber \\
 &+&|10011\rangle +|10101\rangle -|11010\rangle +|11100\rangle).
\end{eqnarray}
The Hoggar and $\psi(\alpha)$ states are provided in the main text in (\ref{eq:hoggar}) and (\ref{eq:psi_alpha}) respectively. \newpage
\bibliographystyle{apsrev4-2}
\bibliography{ref.bib}

\begin{thebibliography}{78}%
\makeatletter
\providecommand \@ifxundefined [1]{%
 \@ifx{#1\undefined}
}%
\providecommand \@ifnum [1]{%
 \ifnum #1\expandafter \@firstoftwo
 \else \expandafter \@secondoftwo
 \fi
}%
\providecommand \@ifx [1]{%
 \ifx #1\expandafter \@firstoftwo
 \else \expandafter \@secondoftwo
 \fi
}%
\providecommand \natexlab [1]{#1}%
\providecommand \enquote  [1]{``#1''}%
\providecommand \bibnamefont  [1]{#1}%
\providecommand \bibfnamefont [1]{#1}%
\providecommand \citenamefont [1]{#1}%
\providecommand \href@noop [0]{\@secondoftwo}%
\providecommand \href [0]{\begingroup \@sanitize@url \@href}%
\providecommand \@href[1]{\@@startlink{#1}\@@href}%
\providecommand \@@href[1]{\endgroup#1\@@endlink}%
\providecommand \@sanitize@url [0]{\catcode `\\12\catcode `\$12\catcode `\&12\catcode `\#12\catcode `\^12\catcode `\_12\catcode `\%12\relax}%
\providecommand \@@startlink[1]{}%
\providecommand \@@endlink[0]{}%
\providecommand \url  [0]{\begingroup\@sanitize@url \@url }%
\providecommand \@url [1]{\endgroup\@href {#1}{\urlprefix }}%
\providecommand \urlprefix  [0]{URL }%
\providecommand \Eprint [0]{\href }%
\providecommand \doibase [0]{https://doi.org/}%
\providecommand \selectlanguage [0]{\@gobble}%
\providecommand \bibinfo  [0]{\@secondoftwo}%
\providecommand \bibfield  [0]{\@secondoftwo}%
\providecommand \translation [1]{[#1]}%
\providecommand \BibitemOpen [0]{}%
\providecommand \bibitemStop [0]{}%
\providecommand \bibitemNoStop [0]{.\EOS\space}%
\providecommand \EOS [0]{\spacefactor3000\relax}%
\providecommand \BibitemShut  [1]{\csname bibitem#1\endcsname}%
\let\auto@bib@innerbib\@empty
\bibitem [{\citenamefont {Brunner}\ \emph {et~al.}(2014)\citenamefont {Brunner}, \citenamefont {Cavalcanti}, \citenamefont {Pironio}, \citenamefont {Scarani},\ and\ \citenamefont {Wehner}}]{Brunner_2014}%
  \BibitemOpen
  \bibfield  {author} {\bibinfo {author} {\bibfnamefont {N.}~\bibnamefont {Brunner}}, \bibinfo {author} {\bibfnamefont {D.}~\bibnamefont {Cavalcanti}}, \bibinfo {author} {\bibfnamefont {S.}~\bibnamefont {Pironio}}, \bibinfo {author} {\bibfnamefont {V.}~\bibnamefont {Scarani}},\ and\ \bibinfo {author} {\bibfnamefont {S.}~\bibnamefont {Wehner}},\ }\href {https://doi.org/10.1103/RevModPhys.86.419} {\bibfield  {journal} {\bibinfo  {journal} {Rev. Mod. Phys.}\ }\textbf {\bibinfo {volume} {86}},\ \bibinfo {pages} {419} (\bibinfo {year} {2014})}\BibitemShut {NoStop}%
\bibitem [{\citenamefont {Pironio}\ \emph {et~al.}(2010)\citenamefont {Pironio}, \citenamefont {Acín}, \citenamefont {Massar}, \citenamefont {de~la Giroday}, \citenamefont {Matsukevich}, \citenamefont {Maunz}, \citenamefont {Olmschenk}, \citenamefont {Hayes}, \citenamefont {Luo}, \citenamefont {Manning},\ and\ \citenamefont {Monroe}}]{Pironio_2010}%
  \BibitemOpen
  \bibfield  {author} {\bibinfo {author} {\bibfnamefont {S.}~\bibnamefont {Pironio}}, \bibinfo {author} {\bibfnamefont {A.}~\bibnamefont {Acín}}, \bibinfo {author} {\bibfnamefont {S.}~\bibnamefont {Massar}}, \bibinfo {author} {\bibfnamefont {A.~B.}\ \bibnamefont {de~la Giroday}}, \bibinfo {author} {\bibfnamefont {D.~N.}\ \bibnamefont {Matsukevich}}, \bibinfo {author} {\bibfnamefont {P.}~\bibnamefont {Maunz}}, \bibinfo {author} {\bibfnamefont {S.}~\bibnamefont {Olmschenk}}, \bibinfo {author} {\bibfnamefont {D.}~\bibnamefont {Hayes}}, \bibinfo {author} {\bibfnamefont {L.}~\bibnamefont {Luo}}, \bibinfo {author} {\bibfnamefont {T.~A.}\ \bibnamefont {Manning}},\ and\ \bibinfo {author} {\bibfnamefont {C.}~\bibnamefont {Monroe}},\ }\href {https://doi.org/10.1038/nature09008} {\bibfield  {journal} {\bibinfo  {journal} {Nature}\ }\textbf {\bibinfo {volume} {464}},\ \bibinfo {pages} {1021–1024} (\bibinfo {year} {2010})}\BibitemShut {NoStop}%
\bibitem [{\citenamefont {Colbeck}(2011)}]{Colbeck_2011}%
  \BibitemOpen
  \bibfield  {author} {\bibinfo {author} {\bibfnamefont {R.}~\bibnamefont {Colbeck}},\ }\href {https://arxiv.org/abs/0911.3814} {\bibinfo {title} {Quantum {A}nd {R}elativistic {P}rotocols {F}or {S}ecure {M}ulti-{P}arty {C}omputation}} (\bibinfo {year} {2011}),\ \Eprint {https://arxiv.org/abs/0911.3814} {arXiv:0911.3814 [quant-ph]} \BibitemShut {NoStop}%
\bibitem [{\citenamefont {Colbeck}\ and\ \citenamefont {Kent}(2011)}]{Colbeck_2011_2}%
  \BibitemOpen
  \bibfield  {author} {\bibinfo {author} {\bibfnamefont {R.}~\bibnamefont {Colbeck}}\ and\ \bibinfo {author} {\bibfnamefont {A.}~\bibnamefont {Kent}},\ }\href {https://doi.org/10.1088/1751-8113/44/9/095305} {\bibfield  {journal} {\bibinfo  {journal} {Journal of Physics A: Mathematical and Theoretical}\ }\textbf {\bibinfo {volume} {44}},\ \bibinfo {pages} {095305} (\bibinfo {year} {2011})}\BibitemShut {NoStop}%
\bibitem [{\citenamefont {Arnon-Friedman}\ \emph {et~al.}(2018)\citenamefont {Arnon-Friedman}, \citenamefont {Dupuis}, \citenamefont {Fawzi}, \citenamefont {Renner},\ and\ \citenamefont {Vidick}}]{ArnonFriedman_2018}%
  \BibitemOpen
  \bibfield  {author} {\bibinfo {author} {\bibfnamefont {R.}~\bibnamefont {Arnon-Friedman}}, \bibinfo {author} {\bibfnamefont {F.}~\bibnamefont {Dupuis}}, \bibinfo {author} {\bibfnamefont {O.}~\bibnamefont {Fawzi}}, \bibinfo {author} {\bibfnamefont {R.}~\bibnamefont {Renner}},\ and\ \bibinfo {author} {\bibfnamefont {T.}~\bibnamefont {Vidick}},\ }\bibfield  {journal} {\bibinfo  {journal} {Nature Communications}\ }\textbf {\bibinfo {volume} {9}},\ \href {https://doi.org/10.1038/s41467-017-02307-4} {10.1038/s41467-017-02307-4} (\bibinfo {year} {2018})\BibitemShut {NoStop}%
\bibitem [{\citenamefont {Ekert}(1991)}]{Ekert_1991}%
  \BibitemOpen
  \bibfield  {author} {\bibinfo {author} {\bibfnamefont {A.~K.}\ \bibnamefont {Ekert}},\ }\href {https://doi.org/10.1103/PhysRevLett.67.661} {\bibfield  {journal} {\bibinfo  {journal} {Phys. Rev. Lett.}\ }\textbf {\bibinfo {volume} {67}},\ \bibinfo {pages} {661} (\bibinfo {year} {1991})}\BibitemShut {NoStop}%
\bibitem [{\citenamefont {Mayers}\ and\ \citenamefont {Yao}(1998)}]{Mayers_1998}%
  \BibitemOpen
  \bibfield  {author} {\bibinfo {author} {\bibfnamefont {D.}~\bibnamefont {Mayers}}\ and\ \bibinfo {author} {\bibfnamefont {A.}~\bibnamefont {Yao}},\ }in\ \href {https://doi.org/10.1109/sfcs.1998.743501} {\emph {\bibinfo {booktitle} {Proceedings 39th Annual Symposium on Foundations of Computer Science (Cat. No.98CB36280)}}},\ \bibinfo {series and number} {SFCS-98}\ (\bibinfo  {publisher} {IEEE Comput. Soc},\ \bibinfo {year} {1998})\ p.\ \bibinfo {pages} {503–509}\BibitemShut {NoStop}%
\bibitem [{\citenamefont {Barrett}\ \emph {et~al.}(2005)\citenamefont {Barrett}, \citenamefont {Hardy},\ and\ \citenamefont {Kent}}]{Barrett_2005}%
  \BibitemOpen
  \bibfield  {author} {\bibinfo {author} {\bibfnamefont {J.}~\bibnamefont {Barrett}}, \bibinfo {author} {\bibfnamefont {L.}~\bibnamefont {Hardy}},\ and\ \bibinfo {author} {\bibfnamefont {A.}~\bibnamefont {Kent}},\ }\href {https://doi.org/10.1103/PhysRevLett.95.010503} {\bibfield  {journal} {\bibinfo  {journal} {Phys. Rev. Lett.}\ }\textbf {\bibinfo {volume} {95}},\ \bibinfo {pages} {010503} (\bibinfo {year} {2005})}\BibitemShut {NoStop}%
\bibitem [{\citenamefont {Ac\'{\i}n}\ \emph {et~al.}(2007)\citenamefont {Ac\'{\i}n}, \citenamefont {Brunner}, \citenamefont {Gisin}, \citenamefont {Massar}, \citenamefont {Pironio},\ and\ \citenamefont {Scarani}}]{Acin_2007}%
  \BibitemOpen
  \bibfield  {author} {\bibinfo {author} {\bibfnamefont {A.}~\bibnamefont {Ac\'{\i}n}}, \bibinfo {author} {\bibfnamefont {N.}~\bibnamefont {Brunner}}, \bibinfo {author} {\bibfnamefont {N.}~\bibnamefont {Gisin}}, \bibinfo {author} {\bibfnamefont {S.}~\bibnamefont {Massar}}, \bibinfo {author} {\bibfnamefont {S.}~\bibnamefont {Pironio}},\ and\ \bibinfo {author} {\bibfnamefont {V.}~\bibnamefont {Scarani}},\ }\href {https://doi.org/10.1103/PhysRevLett.98.230501} {\bibfield  {journal} {\bibinfo  {journal} {Phys. Rev. Lett.}\ }\textbf {\bibinfo {volume} {98}},\ \bibinfo {pages} {230501} (\bibinfo {year} {2007})}\BibitemShut {NoStop}%
\bibitem [{\citenamefont {Pironio}\ \emph {et~al.}(2009)\citenamefont {Pironio}, \citenamefont {Acín}, \citenamefont {Brunner}, \citenamefont {Gisin}, \citenamefont {Massar},\ and\ \citenamefont {Scarani}}]{Pironio_2009}%
  \BibitemOpen
  \bibfield  {author} {\bibinfo {author} {\bibfnamefont {S.}~\bibnamefont {Pironio}}, \bibinfo {author} {\bibfnamefont {A.}~\bibnamefont {Acín}}, \bibinfo {author} {\bibfnamefont {N.}~\bibnamefont {Brunner}}, \bibinfo {author} {\bibfnamefont {N.}~\bibnamefont {Gisin}}, \bibinfo {author} {\bibfnamefont {S.}~\bibnamefont {Massar}},\ and\ \bibinfo {author} {\bibfnamefont {V.}~\bibnamefont {Scarani}},\ }\href {https://doi.org/10.1088/1367-2630/11/4/045021} {\bibfield  {journal} {\bibinfo  {journal} {New Journal of Physics}\ }\textbf {\bibinfo {volume} {11}},\ \bibinfo {pages} {045021} (\bibinfo {year} {2009})}\BibitemShut {NoStop}%
\bibitem [{\citenamefont {Ghoreishi}\ \emph {et~al.}(2025)\citenamefont {Ghoreishi}, \citenamefont {Scala}, \citenamefont {Renner}, \citenamefont {Tacca}, \citenamefont {Bouda}, \citenamefont {Walborn},\ and\ \citenamefont {Pawłowski}}]{Pawlowski_review_2025}%
  \BibitemOpen
  \bibfield  {author} {\bibinfo {author} {\bibfnamefont {S.~A.}\ \bibnamefont {Ghoreishi}}, \bibinfo {author} {\bibfnamefont {G.}~\bibnamefont {Scala}}, \bibinfo {author} {\bibfnamefont {R.}~\bibnamefont {Renner}}, \bibinfo {author} {\bibfnamefont {L.~L.}\ \bibnamefont {Tacca}}, \bibinfo {author} {\bibfnamefont {J.}~\bibnamefont {Bouda}}, \bibinfo {author} {\bibfnamefont {S.~P.}\ \bibnamefont {Walborn}},\ and\ \bibinfo {author} {\bibfnamefont {M.}~\bibnamefont {Pawłowski}},\ }\href {https://doi.org/https://doi.org/10.1016/j.physrep.2025.09.006} {\bibfield  {journal} {\bibinfo  {journal} {Physics Reports}\ }\textbf {\bibinfo {volume} {1149}},\ \bibinfo {pages} {1} (\bibinfo {year} {2025})}\BibitemShut {NoStop}%
\bibitem [{\citenamefont {Cleve}\ and\ \citenamefont {Buhrman}(1997)}]{Cleve_1997}%
  \BibitemOpen
  \bibfield  {author} {\bibinfo {author} {\bibfnamefont {R.}~\bibnamefont {Cleve}}\ and\ \bibinfo {author} {\bibfnamefont {H.}~\bibnamefont {Buhrman}},\ }\href {https://doi.org/10.1103/PhysRevA.56.1201} {\bibfield  {journal} {\bibinfo  {journal} {Phys. Rev. A}\ }\textbf {\bibinfo {volume} {56}},\ \bibinfo {pages} {1201} (\bibinfo {year} {1997})}\BibitemShut {NoStop}%
\bibitem [{\citenamefont {Brukner}\ \emph {et~al.}(2002)\citenamefont {Brukner}, \citenamefont {\ifmmode~\dot{Z}\else \.{Z}\fi{}ukowski},\ and\ \citenamefont {Zeilinger}}]{Brukner_2002}%
  \BibitemOpen
  \bibfield  {author} {\bibinfo {author} {\bibfnamefont {{\v C}.}~\bibnamefont {Brukner}}, \bibinfo {author} {\bibfnamefont {M.}~\bibnamefont {\ifmmode~\dot{Z}\else \.{Z}\fi{}ukowski}},\ and\ \bibinfo {author} {\bibfnamefont {A.}~\bibnamefont {Zeilinger}},\ }\href {https://doi.org/10.1103/PhysRevLett.89.197901} {\bibfield  {journal} {\bibinfo  {journal} {Phys. Rev. Lett.}\ }\textbf {\bibinfo {volume} {89}},\ \bibinfo {pages} {197901} (\bibinfo {year} {2002})}\BibitemShut {NoStop}%
\bibitem [{\citenamefont {Aolita}\ \emph {et~al.}(2012)\citenamefont {Aolita}, \citenamefont {Gallego}, \citenamefont {Cabello},\ and\ \citenamefont {Ac\'{\i}n}}]{Aolita_2012}%
  \BibitemOpen
  \bibfield  {author} {\bibinfo {author} {\bibfnamefont {L.}~\bibnamefont {Aolita}}, \bibinfo {author} {\bibfnamefont {R.}~\bibnamefont {Gallego}}, \bibinfo {author} {\bibfnamefont {A.}~\bibnamefont {Cabello}},\ and\ \bibinfo {author} {\bibfnamefont {A.}~\bibnamefont {Ac\'{\i}n}},\ }\href {https://doi.org/10.1103/PhysRevLett.108.100401} {\bibfield  {journal} {\bibinfo  {journal} {Phys. Rev. Lett.}\ }\textbf {\bibinfo {volume} {108}},\ \bibinfo {pages} {100401} (\bibinfo {year} {2012})}\BibitemShut {NoStop}%
\bibitem [{\citenamefont {Moreno}\ \emph {et~al.}(2020)\citenamefont {Moreno}, \citenamefont {Brito}, \citenamefont {Nery},\ and\ \citenamefont {Chaves}}]{Moreno_2020}%
  \BibitemOpen
  \bibfield  {author} {\bibinfo {author} {\bibfnamefont {M.~G.~M.}\ \bibnamefont {Moreno}}, \bibinfo {author} {\bibfnamefont {S.}~\bibnamefont {Brito}}, \bibinfo {author} {\bibfnamefont {R.~V.}\ \bibnamefont {Nery}},\ and\ \bibinfo {author} {\bibfnamefont {R.}~\bibnamefont {Chaves}},\ }\href {https://doi.org/10.1103/PhysRevA.101.052339} {\bibfield  {journal} {\bibinfo  {journal} {Phys. Rev. A}\ }\textbf {\bibinfo {volume} {101}},\ \bibinfo {pages} {052339} (\bibinfo {year} {2020})}\BibitemShut {NoStop}%
\bibitem [{\citenamefont {Aspect}\ \emph {et~al.}(1982)\citenamefont {Aspect}, \citenamefont {Dalibard},\ and\ \citenamefont {Roger}}]{Aspect1982b}%
  \BibitemOpen
  \bibfield  {author} {\bibinfo {author} {\bibfnamefont {A.}~\bibnamefont {Aspect}}, \bibinfo {author} {\bibfnamefont {J.}~\bibnamefont {Dalibard}},\ and\ \bibinfo {author} {\bibfnamefont {G.}~\bibnamefont {Roger}},\ }\href@noop {} {\bibfield  {journal} {\bibinfo  {journal} {Phys. Rev. Lett.}\ }\textbf {\bibinfo {volume} {49}},\ \bibinfo {pages} {1804} (\bibinfo {year} {1982})}\BibitemShut {NoStop}%
\bibitem [{\citenamefont {Weihs}\ \emph {et~al.}(1998)\citenamefont {Weihs}, \citenamefont {Jennewein}, \citenamefont {Simon}, \citenamefont {Weinfurter},\ and\ \citenamefont {Zeilinger}}]{Weihs1998}%
  \BibitemOpen
  \bibfield  {author} {\bibinfo {author} {\bibfnamefont {G.}~\bibnamefont {Weihs}}, \bibinfo {author} {\bibfnamefont {T.}~\bibnamefont {Jennewein}}, \bibinfo {author} {\bibfnamefont {C.}~\bibnamefont {Simon}}, \bibinfo {author} {\bibfnamefont {H.}~\bibnamefont {Weinfurter}},\ and\ \bibinfo {author} {\bibfnamefont {A.}~\bibnamefont {Zeilinger}},\ }\href@noop {} {\bibfield  {journal} {\bibinfo  {journal} {Phys. Rev. Lett.}\ }\textbf {\bibinfo {volume} {81}},\ \bibinfo {pages} {5039} (\bibinfo {year} {1998})}\BibitemShut {NoStop}%
\bibitem [{\citenamefont {Rowe}\ \emph {et~al.}(2001)\citenamefont {Rowe}, \citenamefont {Kielpinski}, \citenamefont {Meyer}, \citenamefont {Sackett}, \citenamefont {Itano}, \citenamefont {Monroe},\ and\ \citenamefont {Wineland}}]{Rowe2001}%
  \BibitemOpen
  \bibfield  {author} {\bibinfo {author} {\bibfnamefont {M.~A.}\ \bibnamefont {Rowe}}, \bibinfo {author} {\bibfnamefont {D.}~\bibnamefont {Kielpinski}}, \bibinfo {author} {\bibfnamefont {V.}~\bibnamefont {Meyer}}, \bibinfo {author} {\bibfnamefont {C.~A.}\ \bibnamefont {Sackett}}, \bibinfo {author} {\bibfnamefont {W.~M.}\ \bibnamefont {Itano}}, \bibinfo {author} {\bibfnamefont {C.}~\bibnamefont {Monroe}},\ and\ \bibinfo {author} {\bibfnamefont {D.~J.}\ \bibnamefont {Wineland}},\ }\href@noop {} {\bibfield  {journal} {\bibinfo  {journal} {Nature}\ }\textbf {\bibinfo {volume} {409}},\ \bibinfo {pages} {791–794} (\bibinfo {year} {2001})}\BibitemShut {NoStop}%
\bibitem [{\citenamefont {Pan}\ \emph {et~al.}(2000)\citenamefont {Pan}, \citenamefont {Bouwmeester}, \citenamefont {Daniell}, \citenamefont {Weinfurter},\ and\ \citenamefont {Zeilinger}}]{Pan2000}%
  \BibitemOpen
  \bibfield  {author} {\bibinfo {author} {\bibfnamefont {J.-W.}\ \bibnamefont {Pan}}, \bibinfo {author} {\bibfnamefont {D.}~\bibnamefont {Bouwmeester}}, \bibinfo {author} {\bibfnamefont {M.}~\bibnamefont {Daniell}}, \bibinfo {author} {\bibfnamefont {H.}~\bibnamefont {Weinfurter}},\ and\ \bibinfo {author} {\bibfnamefont {A.}~\bibnamefont {Zeilinger}},\ }\href@noop {} {\bibfield  {journal} {\bibinfo  {journal} {Nature}\ }\textbf {\bibinfo {volume} {403}},\ \bibinfo {pages} {515} (\bibinfo {year} {2000})}\BibitemShut {NoStop}%
\bibitem [{\citenamefont {Walther}\ \emph {et~al.}(2005)\citenamefont {Walther}, \citenamefont {Aspelmeyer}, \citenamefont {Resch},\ and\ \citenamefont {Zeilinger}}]{Walther2005}%
  \BibitemOpen
  \bibfield  {author} {\bibinfo {author} {\bibfnamefont {P.}~\bibnamefont {Walther}}, \bibinfo {author} {\bibfnamefont {M.}~\bibnamefont {Aspelmeyer}}, \bibinfo {author} {\bibfnamefont {K.~J.}\ \bibnamefont {Resch}},\ and\ \bibinfo {author} {\bibfnamefont {A.}~\bibnamefont {Zeilinger}},\ }\href@noop {} {\bibfield  {journal} {\bibinfo  {journal} {Phys. Rev. Lett.}\ }\textbf {\bibinfo {volume} {95}},\ \bibinfo {pages} {020403} (\bibinfo {year} {2005})}\BibitemShut {NoStop}%
\bibitem [{\citenamefont {Hensen}\ \emph {et~al.}(2015)\citenamefont {Hensen}, \citenamefont {Bernien}, \citenamefont {Dréau}, \citenamefont {Reiserer}, \citenamefont {Kalb}, \citenamefont {Blok}, \citenamefont {Ruitenberg}, \citenamefont {Vermeulen}, \citenamefont {Schouten}, \citenamefont {Abell{\'a}n}, \citenamefont {Amaya}, \citenamefont {Pruneri}, \citenamefont {Mitchell}, \citenamefont {Markham}, \citenamefont {Twitchen}, \citenamefont {Elkouss}, \citenamefont {Wehner}, \citenamefont {Taminiau},\ and\ \citenamefont {Hanson}}]{Hensen2015}%
  \BibitemOpen
  \bibfield  {author} {\bibinfo {author} {\bibfnamefont {B.}~\bibnamefont {Hensen}}, \bibinfo {author} {\bibfnamefont {H.}~\bibnamefont {Bernien}}, \bibinfo {author} {\bibfnamefont {A.~E.}\ \bibnamefont {Dréau}}, \bibinfo {author} {\bibfnamefont {A.}~\bibnamefont {Reiserer}}, \bibinfo {author} {\bibfnamefont {N.}~\bibnamefont {Kalb}}, \bibinfo {author} {\bibfnamefont {M.~S.}\ \bibnamefont {Blok}}, \bibinfo {author} {\bibfnamefont {J.}~\bibnamefont {Ruitenberg}}, \bibinfo {author} {\bibfnamefont {R.~F.~L.}\ \bibnamefont {Vermeulen}}, \bibinfo {author} {\bibfnamefont {R.~N.}\ \bibnamefont {Schouten}}, \bibinfo {author} {\bibfnamefont {C.}~\bibnamefont {Abell{\'a}n}}, \bibinfo {author} {\bibfnamefont {W.}~\bibnamefont {Amaya}}, \bibinfo {author} {\bibfnamefont {V.}~\bibnamefont {Pruneri}}, \bibinfo {author} {\bibfnamefont {M.~W.}\ \bibnamefont {Mitchell}}, \bibinfo {author} {\bibfnamefont {M.}~\bibnamefont {Markham}}, \bibinfo {author} {\bibfnamefont {D.~J.}\ \bibnamefont {Twitchen}}, \bibinfo {author}
  {\bibfnamefont {D.}~\bibnamefont {Elkouss}}, \bibinfo {author} {\bibfnamefont {S.}~\bibnamefont {Wehner}}, \bibinfo {author} {\bibfnamefont {T.~H.}\ \bibnamefont {Taminiau}},\ and\ \bibinfo {author} {\bibfnamefont {R.}~\bibnamefont {Hanson}},\ }\href@noop {} {\bibfield  {journal} {\bibinfo  {journal} {Nature}\ }\textbf {\bibinfo {volume} {526}},\ \bibinfo {pages} {682–686} (\bibinfo {year} {2015})}\BibitemShut {NoStop}%
\bibitem [{\citenamefont {Giustina}\ \emph {et~al.}(2015)\citenamefont {Giustina}, \citenamefont {Versteegh}, \citenamefont {Wengerowsky}, \citenamefont {Handsteiner}, \citenamefont {Hochrainer}, \citenamefont {Phelan}, \citenamefont {Steinlechner}, \citenamefont {Kofler}, \citenamefont {Larsson}, \citenamefont {Abell{\'a}n}, \citenamefont {Amaya}, \citenamefont {Pruneri}, \citenamefont {Mitchell}, \citenamefont {Beyer}, \citenamefont {Gerrits}, \citenamefont {Lita}, \citenamefont {Shalm}, \citenamefont {Nam}, \citenamefont {Scheidl}, \citenamefont {Ursin}, \citenamefont {Wittmann},\ and\ \citenamefont {Zeilinger}}]{Giustina2015}%
  \BibitemOpen
  \bibfield  {author} {\bibinfo {author} {\bibfnamefont {M.}~\bibnamefont {Giustina}}, \bibinfo {author} {\bibfnamefont {M.~A.~M.}\ \bibnamefont {Versteegh}}, \bibinfo {author} {\bibfnamefont {S.}~\bibnamefont {Wengerowsky}}, \bibinfo {author} {\bibfnamefont {J.}~\bibnamefont {Handsteiner}}, \bibinfo {author} {\bibfnamefont {A.}~\bibnamefont {Hochrainer}}, \bibinfo {author} {\bibfnamefont {K.}~\bibnamefont {Phelan}}, \bibinfo {author} {\bibfnamefont {F.}~\bibnamefont {Steinlechner}}, \bibinfo {author} {\bibfnamefont {J.}~\bibnamefont {Kofler}}, \bibinfo {author} {\bibfnamefont {J.-A.}\ \bibnamefont {Larsson}}, \bibinfo {author} {\bibfnamefont {C.}~\bibnamefont {Abell{\'a}n}}, \bibinfo {author} {\bibfnamefont {W.}~\bibnamefont {Amaya}}, \bibinfo {author} {\bibfnamefont {V.}~\bibnamefont {Pruneri}}, \bibinfo {author} {\bibfnamefont {M.~W.}\ \bibnamefont {Mitchell}}, \bibinfo {author} {\bibfnamefont {J.}~\bibnamefont {Beyer}}, \bibinfo {author} {\bibfnamefont {T.}~\bibnamefont {Gerrits}}, \bibinfo {author}
  {\bibfnamefont {A.~E.}\ \bibnamefont {Lita}}, \bibinfo {author} {\bibfnamefont {L.~K.}\ \bibnamefont {Shalm}}, \bibinfo {author} {\bibfnamefont {S.~W.}\ \bibnamefont {Nam}}, \bibinfo {author} {\bibfnamefont {T.}~\bibnamefont {Scheidl}}, \bibinfo {author} {\bibfnamefont {R.}~\bibnamefont {Ursin}}, \bibinfo {author} {\bibfnamefont {B.}~\bibnamefont {Wittmann}},\ and\ \bibinfo {author} {\bibfnamefont {A.}~\bibnamefont {Zeilinger}},\ }\href@noop {} {\bibfield  {journal} {\bibinfo  {journal} {Phys. Rev. Lett.}\ }\textbf {\bibinfo {volume} {115}} (\bibinfo {year} {2015})}\BibitemShut {NoStop}%
\bibitem [{\citenamefont {Shalm}\ \emph {et~al.}(2015)\citenamefont {Shalm}, \citenamefont {Meyer-Scott}, \citenamefont {Christensen}, \citenamefont {Bierhorst}, \citenamefont {Wayne}, \citenamefont {Stevens}, \citenamefont {Gerrits}, \citenamefont {Glancy}, \citenamefont {Hamel}, \citenamefont {Allman}, \citenamefont {Coakley}, \citenamefont {Dyer}, \citenamefont {Hodge}, \citenamefont {Lita}, \citenamefont {Verma}, \citenamefont {Lambrocco}, \citenamefont {Tortorici}, \citenamefont {Migdall}, \citenamefont {Zhang}, \citenamefont {Kumor}, \citenamefont {Farr}, \citenamefont {Marsili}, \citenamefont {Shaw}, \citenamefont {Stern}, \citenamefont {Abell{\'a}n}, \citenamefont {Amaya}, \citenamefont {Pruneri}, \citenamefont {Jennewein}, \citenamefont {Mitchell}, \citenamefont {Kwiat}, \citenamefont {Bienfang}, \citenamefont {Mirin}, \citenamefont {Knill},\ and\ \citenamefont {Nam}}]{Shalm2015}%
  \BibitemOpen
  \bibfield  {author} {\bibinfo {author} {\bibfnamefont {L.~K.}\ \bibnamefont {Shalm}}, \bibinfo {author} {\bibfnamefont {E.}~\bibnamefont {Meyer-Scott}}, \bibinfo {author} {\bibfnamefont {B.~G.}\ \bibnamefont {Christensen}}, \bibinfo {author} {\bibfnamefont {P.}~\bibnamefont {Bierhorst}}, \bibinfo {author} {\bibfnamefont {M.~A.}\ \bibnamefont {Wayne}}, \bibinfo {author} {\bibfnamefont {M.~J.}\ \bibnamefont {Stevens}}, \bibinfo {author} {\bibfnamefont {T.}~\bibnamefont {Gerrits}}, \bibinfo {author} {\bibfnamefont {S.}~\bibnamefont {Glancy}}, \bibinfo {author} {\bibfnamefont {D.~R.}\ \bibnamefont {Hamel}}, \bibinfo {author} {\bibfnamefont {M.~S.}\ \bibnamefont {Allman}}, \bibinfo {author} {\bibfnamefont {K.~J.}\ \bibnamefont {Coakley}}, \bibinfo {author} {\bibfnamefont {S.~D.}\ \bibnamefont {Dyer}}, \bibinfo {author} {\bibfnamefont {C.}~\bibnamefont {Hodge}}, \bibinfo {author} {\bibfnamefont {A.~E.}\ \bibnamefont {Lita}}, \bibinfo {author} {\bibfnamefont {V.~B.}\ \bibnamefont {Verma}}, \bibinfo {author}
  {\bibfnamefont {C.}~\bibnamefont {Lambrocco}}, \bibinfo {author} {\bibfnamefont {E.}~\bibnamefont {Tortorici}}, \bibinfo {author} {\bibfnamefont {A.~L.}\ \bibnamefont {Migdall}}, \bibinfo {author} {\bibfnamefont {Y.}~\bibnamefont {Zhang}}, \bibinfo {author} {\bibfnamefont {D.~R.}\ \bibnamefont {Kumor}}, \bibinfo {author} {\bibfnamefont {W.~H.}\ \bibnamefont {Farr}}, \bibinfo {author} {\bibfnamefont {F.}~\bibnamefont {Marsili}}, \bibinfo {author} {\bibfnamefont {M.~D.}\ \bibnamefont {Shaw}}, \bibinfo {author} {\bibfnamefont {J.~A.}\ \bibnamefont {Stern}}, \bibinfo {author} {\bibfnamefont {C.}~\bibnamefont {Abell{\'a}n}}, \bibinfo {author} {\bibfnamefont {W.}~\bibnamefont {Amaya}}, \bibinfo {author} {\bibfnamefont {V.}~\bibnamefont {Pruneri}}, \bibinfo {author} {\bibfnamefont {T.}~\bibnamefont {Jennewein}}, \bibinfo {author} {\bibfnamefont {M.~W.}\ \bibnamefont {Mitchell}}, \bibinfo {author} {\bibfnamefont {P.~G.}\ \bibnamefont {Kwiat}}, \bibinfo {author} {\bibfnamefont {J.~C.}\ \bibnamefont {Bienfang}},
  \bibinfo {author} {\bibfnamefont {R.~P.}\ \bibnamefont {Mirin}}, \bibinfo {author} {\bibfnamefont {E.}~\bibnamefont {Knill}},\ and\ \bibinfo {author} {\bibfnamefont {S.~W.}\ \bibnamefont {Nam}},\ }\href@noop {} {\bibfield  {journal} {\bibinfo  {journal} {Phys. Rev. Lett.}\ }\textbf {\bibinfo {volume} {115}} (\bibinfo {year} {2015})}\BibitemShut {NoStop}%
\bibitem [{\citenamefont {Rosenfeld}\ \emph {et~al.}(2017)\citenamefont {Rosenfeld}, \citenamefont {Burchardt}, \citenamefont {Garthoff}, \citenamefont {Redeker}, \citenamefont {Ortegel}, \citenamefont {Rau},\ and\ \citenamefont {Weinfurter}}]{Rosenfeld2017}%
  \BibitemOpen
  \bibfield  {author} {\bibinfo {author} {\bibfnamefont {W.}~\bibnamefont {Rosenfeld}}, \bibinfo {author} {\bibfnamefont {D.}~\bibnamefont {Burchardt}}, \bibinfo {author} {\bibfnamefont {R.}~\bibnamefont {Garthoff}}, \bibinfo {author} {\bibfnamefont {K.}~\bibnamefont {Redeker}}, \bibinfo {author} {\bibfnamefont {N.}~\bibnamefont {Ortegel}}, \bibinfo {author} {\bibfnamefont {M.}~\bibnamefont {Rau}},\ and\ \bibinfo {author} {\bibfnamefont {H.}~\bibnamefont {Weinfurter}},\ }\href@noop {} {\bibfield  {journal} {\bibinfo  {journal} {Phys. Rev. Lett.}\ }\textbf {\bibinfo {volume} {119}} (\bibinfo {year} {2017})}\BibitemShut {NoStop}%
\bibitem [{\citenamefont {Rauch}\ \emph {et~al.}(2018)\citenamefont {Rauch}, \citenamefont {Handsteiner}, \citenamefont {Hochrainer}, \citenamefont {Gallicchio}, \citenamefont {Friedman}, \citenamefont {Leung}, \citenamefont {Liu}, \citenamefont {Bulla}, \citenamefont {Ecker}, \citenamefont {Steinlechner}, \citenamefont {Ursin}, \citenamefont {Hu}, \citenamefont {Leon}, \citenamefont {Benn}, \citenamefont {Ghedina}, \citenamefont {Cecconi}, \citenamefont {Guth}, \citenamefont {Kaiser}, \citenamefont {Scheidl},\ and\ \citenamefont {Zeilinger}}]{Rauch2018}%
  \BibitemOpen
  \bibfield  {author} {\bibinfo {author} {\bibfnamefont {D.}~\bibnamefont {Rauch}}, \bibinfo {author} {\bibfnamefont {J.}~\bibnamefont {Handsteiner}}, \bibinfo {author} {\bibfnamefont {A.}~\bibnamefont {Hochrainer}}, \bibinfo {author} {\bibfnamefont {J.}~\bibnamefont {Gallicchio}}, \bibinfo {author} {\bibfnamefont {A.~S.}\ \bibnamefont {Friedman}}, \bibinfo {author} {\bibfnamefont {C.}~\bibnamefont {Leung}}, \bibinfo {author} {\bibfnamefont {B.}~\bibnamefont {Liu}}, \bibinfo {author} {\bibfnamefont {L.}~\bibnamefont {Bulla}}, \bibinfo {author} {\bibfnamefont {S.}~\bibnamefont {Ecker}}, \bibinfo {author} {\bibfnamefont {F.}~\bibnamefont {Steinlechner}}, \bibinfo {author} {\bibfnamefont {R.}~\bibnamefont {Ursin}}, \bibinfo {author} {\bibfnamefont {B.}~\bibnamefont {Hu}}, \bibinfo {author} {\bibfnamefont {D.}~\bibnamefont {Leon}}, \bibinfo {author} {\bibfnamefont {C.}~\bibnamefont {Benn}}, \bibinfo {author} {\bibfnamefont {A.}~\bibnamefont {Ghedina}}, \bibinfo {author} {\bibfnamefont {M.}~\bibnamefont
  {Cecconi}}, \bibinfo {author} {\bibfnamefont {A.~H.}\ \bibnamefont {Guth}}, \bibinfo {author} {\bibfnamefont {D.~I.}\ \bibnamefont {Kaiser}}, \bibinfo {author} {\bibfnamefont {T.}~\bibnamefont {Scheidl}},\ and\ \bibinfo {author} {\bibfnamefont {A.}~\bibnamefont {Zeilinger}},\ }\href@noop {} {\bibfield  {journal} {\bibinfo  {journal} {Phys. Rev. Lett.}\ }\textbf {\bibinfo {volume} {121}} (\bibinfo {year} {2018})}\BibitemShut {NoStop}%
\bibitem [{\citenamefont {Storz}\ \emph {et~al.}(2023)\citenamefont {Storz}, \citenamefont {Sch\"{a}r}, \citenamefont {Kulikov}, \citenamefont {Magnard}, \citenamefont {Kurpiers}, \citenamefont {L\"{u}tolf}, \citenamefont {Walter}, \citenamefont {Copetudo}, \citenamefont {Reuer}, \citenamefont {Akin}, \citenamefont {Besse}, \citenamefont {Gabureac}, \citenamefont {Norris}, \citenamefont {Rosario}, \citenamefont {Martin}, \citenamefont {Martinez}, \citenamefont {Amaya}, \citenamefont {Mitchell}, \citenamefont {Abellan}, \citenamefont {Bancal}, \citenamefont {Sangouard}, \citenamefont {Royer}, \citenamefont {Blais},\ and\ \citenamefont {Wallraff}}]{Storz2023}%
  \BibitemOpen
  \bibfield  {author} {\bibinfo {author} {\bibfnamefont {S.}~\bibnamefont {Storz}}, \bibinfo {author} {\bibfnamefont {J.}~\bibnamefont {Sch\"{a}r}}, \bibinfo {author} {\bibfnamefont {A.}~\bibnamefont {Kulikov}}, \bibinfo {author} {\bibfnamefont {P.}~\bibnamefont {Magnard}}, \bibinfo {author} {\bibfnamefont {P.}~\bibnamefont {Kurpiers}}, \bibinfo {author} {\bibfnamefont {J.}~\bibnamefont {L\"{u}tolf}}, \bibinfo {author} {\bibfnamefont {T.}~\bibnamefont {Walter}}, \bibinfo {author} {\bibfnamefont {A.}~\bibnamefont {Copetudo}}, \bibinfo {author} {\bibfnamefont {K.}~\bibnamefont {Reuer}}, \bibinfo {author} {\bibfnamefont {A.}~\bibnamefont {Akin}}, \bibinfo {author} {\bibfnamefont {J.-C.}\ \bibnamefont {Besse}}, \bibinfo {author} {\bibfnamefont {M.}~\bibnamefont {Gabureac}}, \bibinfo {author} {\bibfnamefont {G.~J.}\ \bibnamefont {Norris}}, \bibinfo {author} {\bibfnamefont {A.}~\bibnamefont {Rosario}}, \bibinfo {author} {\bibfnamefont {F.}~\bibnamefont {Martin}}, \bibinfo {author} {\bibfnamefont {J.}~\bibnamefont
  {Martinez}}, \bibinfo {author} {\bibfnamefont {W.}~\bibnamefont {Amaya}}, \bibinfo {author} {\bibfnamefont {M.~W.}\ \bibnamefont {Mitchell}}, \bibinfo {author} {\bibfnamefont {C.}~\bibnamefont {Abellan}}, \bibinfo {author} {\bibfnamefont {J.-D.}\ \bibnamefont {Bancal}}, \bibinfo {author} {\bibfnamefont {N.}~\bibnamefont {Sangouard}}, \bibinfo {author} {\bibfnamefont {B.}~\bibnamefont {Royer}}, \bibinfo {author} {\bibfnamefont {A.}~\bibnamefont {Blais}},\ and\ \bibinfo {author} {\bibfnamefont {A.}~\bibnamefont {Wallraff}},\ }\href {https://doi.org/10.1038/s41586-023-05885-0} {\bibfield  {journal} {\bibinfo  {journal} {Nature}\ }\textbf {\bibinfo {volume} {617}},\ \bibinfo {pages} {265–270} (\bibinfo {year} {2023})}\BibitemShut {NoStop}%
\bibitem [{\citenamefont {\ifmmode~\dot{Z}\else \.{Z}\fi{}ukowski}\ and\ \citenamefont {Brukner}(2002)}]{Zukowski_2002}%
  \BibitemOpen
  \bibfield  {author} {\bibinfo {author} {\bibfnamefont {M.}~\bibnamefont {\ifmmode~\dot{Z}\else \.{Z}\fi{}ukowski}}\ and\ \bibinfo {author} {\bibfnamefont {i.~c.~v.}\ \bibnamefont {Brukner}},\ }\href {https://doi.org/10.1103/PhysRevLett.88.210401} {\bibfield  {journal} {\bibinfo  {journal} {Phys. Rev. Lett.}\ }\textbf {\bibinfo {volume} {88}},\ \bibinfo {pages} {210401} (\bibinfo {year} {2002})}\BibitemShut {NoStop}%
\bibitem [{\citenamefont {Howard}(2015)}]{Howard_2015}%
  \BibitemOpen
  \bibfield  {author} {\bibinfo {author} {\bibfnamefont {M.}~\bibnamefont {Howard}},\ }\href {https://doi.org/10.1103/PhysRevA.91.042103} {\bibfield  {journal} {\bibinfo  {journal} {Phys. Rev. A}\ }\textbf {\bibinfo {volume} {91}},\ \bibinfo {pages} {042103} (\bibinfo {year} {2015})}\BibitemShut {NoStop}%
\bibitem [{\citenamefont {Clauser}\ \emph {et~al.}(1969)\citenamefont {Clauser}, \citenamefont {Horne}, \citenamefont {Shimony},\ and\ \citenamefont {Holt}}]{Clauser_1969}%
  \BibitemOpen
  \bibfield  {author} {\bibinfo {author} {\bibfnamefont {J.~F.}\ \bibnamefont {Clauser}}, \bibinfo {author} {\bibfnamefont {M.~A.}\ \bibnamefont {Horne}}, \bibinfo {author} {\bibfnamefont {A.}~\bibnamefont {Shimony}},\ and\ \bibinfo {author} {\bibfnamefont {R.~A.}\ \bibnamefont {Holt}},\ }\href {https://doi.org/10.1103/PhysRevLett.23.880} {\bibfield  {journal} {\bibinfo  {journal} {Phys. Rev. Lett.}\ }\textbf {\bibinfo {volume} {23}},\ \bibinfo {pages} {880} (\bibinfo {year} {1969})}\BibitemShut {NoStop}%
\bibitem [{\citenamefont {Mermin}(1990)}]{MERMIN}%
  \BibitemOpen
  \bibfield  {author} {\bibinfo {author} {\bibfnamefont {N.~D.}\ \bibnamefont {Mermin}},\ }\href@noop {} {\bibfield  {journal} {\bibinfo  {journal} {Phys. Rev. Lett.}\ }\textbf {\bibinfo {volume} {65}},\ \bibinfo {pages} {1838} (\bibinfo {year} {1990})}\BibitemShut {NoStop}%
\bibitem [{\citenamefont {Ardehali}(1992)}]{Ardehali_1992}%
  \BibitemOpen
  \bibfield  {author} {\bibinfo {author} {\bibfnamefont {M.}~\bibnamefont {Ardehali}},\ }\href@noop {} {\bibfield  {journal} {\bibinfo  {journal} {Phys. Rev. A}\ }\textbf {\bibinfo {volume} {46}},\ \bibinfo {pages} {5375} (\bibinfo {year} {1992})}\BibitemShut {NoStop}%
\bibitem [{\citenamefont {Belinskiĭ}\ and\ \citenamefont {Klyshko}(1993)}]{Belinskii1993}%
  \BibitemOpen
  \bibfield  {author} {\bibinfo {author} {\bibfnamefont {A.~V.}\ \bibnamefont {Belinskiĭ}}\ and\ \bibinfo {author} {\bibfnamefont {D.~N.}\ \bibnamefont {Klyshko}},\ }\href@noop {} {\bibfield  {journal} {\bibinfo  {journal} {Phys.-Uspekhi}\ }\textbf {\bibinfo {volume} {36}},\ \bibinfo {pages} {653} (\bibinfo {year} {1993})}\BibitemShut {NoStop}%
\bibitem [{\citenamefont {G\"uhne}\ \emph {et~al.}(2005)\citenamefont {G\"uhne}, \citenamefont {T\'oth}, \citenamefont {Hyllus},\ and\ \citenamefont {Briegel}}]{Guhne_2005}%
  \BibitemOpen
  \bibfield  {author} {\bibinfo {author} {\bibfnamefont {O.}~\bibnamefont {G\"uhne}}, \bibinfo {author} {\bibfnamefont {G.}~\bibnamefont {T\'oth}}, \bibinfo {author} {\bibfnamefont {P.}~\bibnamefont {Hyllus}},\ and\ \bibinfo {author} {\bibfnamefont {H.~J.}\ \bibnamefont {Briegel}},\ }\href {https://doi.org/10.1103/PhysRevLett.95.120405} {\bibfield  {journal} {\bibinfo  {journal} {Phys. Rev. Lett.}\ }\textbf {\bibinfo {volume} {95}},\ \bibinfo {pages} {120405} (\bibinfo {year} {2005})}\BibitemShut {NoStop}%
\bibitem [{\citenamefont {Scarani}\ \emph {et~al.}(2005)\citenamefont {Scarani}, \citenamefont {Ac\'{\i}n}, \citenamefont {Schenck},\ and\ \citenamefont {Aspelmeyer}}]{Scarani_2005}%
  \BibitemOpen
  \bibfield  {author} {\bibinfo {author} {\bibfnamefont {V.}~\bibnamefont {Scarani}}, \bibinfo {author} {\bibfnamefont {A.}~\bibnamefont {Ac\'{\i}n}}, \bibinfo {author} {\bibfnamefont {E.}~\bibnamefont {Schenck}},\ and\ \bibinfo {author} {\bibfnamefont {M.}~\bibnamefont {Aspelmeyer}},\ }\href {https://doi.org/10.1103/PhysRevA.71.042325} {\bibfield  {journal} {\bibinfo  {journal} {Phys. Rev. A}\ }\textbf {\bibinfo {volume} {71}},\ \bibinfo {pages} {042325} (\bibinfo {year} {2005})}\BibitemShut {NoStop}%
\bibitem [{\citenamefont {T\'oth}\ \emph {et~al.}(2006)\citenamefont {T\'oth}, \citenamefont {G\"uhne},\ and\ \citenamefont {Briegel}}]{Toth_2006}%
  \BibitemOpen
  \bibfield  {author} {\bibinfo {author} {\bibfnamefont {G.}~\bibnamefont {T\'oth}}, \bibinfo {author} {\bibfnamefont {O.}~\bibnamefont {G\"uhne}},\ and\ \bibinfo {author} {\bibfnamefont {H.~J.}\ \bibnamefont {Briegel}},\ }\href {https://doi.org/10.1103/PhysRevA.73.022303} {\bibfield  {journal} {\bibinfo  {journal} {Phys. Rev. A}\ }\textbf {\bibinfo {volume} {73}},\ \bibinfo {pages} {022303} (\bibinfo {year} {2006})}\BibitemShut {NoStop}%
\bibitem [{\citenamefont {G\"uhne}\ and\ \citenamefont {Cabello}(2008)}]{Guhne_2008}%
  \BibitemOpen
  \bibfield  {author} {\bibinfo {author} {\bibfnamefont {O.}~\bibnamefont {G\"uhne}}\ and\ \bibinfo {author} {\bibfnamefont {A.}~\bibnamefont {Cabello}},\ }\href {https://doi.org/10.1103/PhysRevA.77.032108} {\bibfield  {journal} {\bibinfo  {journal} {Phys. Rev. A}\ }\textbf {\bibinfo {volume} {77}},\ \bibinfo {pages} {032108} (\bibinfo {year} {2008})}\BibitemShut {NoStop}%
\bibitem [{\citenamefont {Baccari}\ \emph {et~al.}(2020)\citenamefont {Baccari}, \citenamefont {Augusiak}, \citenamefont {\ifmmode \check{S}\else \v{S}\fi{}upi\ifmmode~\acute{c}\else \'{c}\fi{}}, \citenamefont {Tura},\ and\ \citenamefont {Ac\'{\i}n}}]{Baccari_2020}%
  \BibitemOpen
  \bibfield  {author} {\bibinfo {author} {\bibfnamefont {F.}~\bibnamefont {Baccari}}, \bibinfo {author} {\bibfnamefont {R.}~\bibnamefont {Augusiak}}, \bibinfo {author} {\bibfnamefont {I.}~\bibnamefont {\ifmmode \check{S}\else \v{S}\fi{}upi\ifmmode~\acute{c}\else \'{c}\fi{}}}, \bibinfo {author} {\bibfnamefont {J.}~\bibnamefont {Tura}},\ and\ \bibinfo {author} {\bibfnamefont {A.}~\bibnamefont {Ac\'{\i}n}},\ }\href {https://doi.org/10.1103/PhysRevLett.124.020402} {\bibfield  {journal} {\bibinfo  {journal} {Phys. Rev. Lett.}\ }\textbf {\bibinfo {volume} {124}},\ \bibinfo {pages} {020402} (\bibinfo {year} {2020})}\BibitemShut {NoStop}%
\bibitem [{\citenamefont {Gottesman}(1997)}]{Gottesman_1997}%
  \BibitemOpen
  \bibfield  {author} {\bibinfo {author} {\bibfnamefont {D.}~\bibnamefont {Gottesman}},\ }\href {https://doi.org/10.48550/ARXIV.QUANT-PH/9705052} {\bibinfo {title} {Stabilizer codes and quantum error correction}} (\bibinfo {year} {1997})\BibitemShut {NoStop}%
\bibitem [{\citenamefont {Gottesman}(1998)}]{Gottesman_1998}%
  \BibitemOpen
  \bibfield  {author} {\bibinfo {author} {\bibfnamefont {D.}~\bibnamefont {Gottesman}},\ }\href {https://doi.org/10.48550/ARXIV.QUANT-PH/9807006} {\bibinfo {title} {The heisenberg representation of quantum computers}} (\bibinfo {year} {1998})\BibitemShut {NoStop}%
\bibitem [{\citenamefont {Aaronson}\ and\ \citenamefont {Gottesman}(2004)}]{Aaronson_2004}%
  \BibitemOpen
  \bibfield  {author} {\bibinfo {author} {\bibfnamefont {S.}~\bibnamefont {Aaronson}}\ and\ \bibinfo {author} {\bibfnamefont {D.}~\bibnamefont {Gottesman}},\ }\href {https://doi.org/10.1103/PhysRevA.70.052328} {\bibfield  {journal} {\bibinfo  {journal} {Phys. Rev. A}\ }\textbf {\bibinfo {volume} {70}},\ \bibinfo {pages} {052328} (\bibinfo {year} {2004})}\BibitemShut {NoStop}%
\bibitem [{\citenamefont {Howard}\ and\ \citenamefont {Vala}(2012)}]{Howard_2012}%
  \BibitemOpen
  \bibfield  {author} {\bibinfo {author} {\bibfnamefont {M.}~\bibnamefont {Howard}}\ and\ \bibinfo {author} {\bibfnamefont {J.}~\bibnamefont {Vala}},\ }\href {https://doi.org/10.1103/PhysRevA.85.022304} {\bibfield  {journal} {\bibinfo  {journal} {Phys. Rev. A}\ }\textbf {\bibinfo {volume} {85}},\ \bibinfo {pages} {022304} (\bibinfo {year} {2012})}\BibitemShut {NoStop}%
\bibitem [{\citenamefont {Howard}\ \emph {et~al.}(2014)\citenamefont {Howard}, \citenamefont {Wallman}, \citenamefont {Veitch},\ and\ \citenamefont {Emerson}}]{Howard_2014}%
  \BibitemOpen
  \bibfield  {author} {\bibinfo {author} {\bibfnamefont {M.}~\bibnamefont {Howard}}, \bibinfo {author} {\bibfnamefont {J.}~\bibnamefont {Wallman}}, \bibinfo {author} {\bibfnamefont {V.}~\bibnamefont {Veitch}},\ and\ \bibinfo {author} {\bibfnamefont {J.}~\bibnamefont {Emerson}},\ }\href {https://doi.org/10.1038/nature13460} {\bibfield  {journal} {\bibinfo  {journal} {Nature}\ }\textbf {\bibinfo {volume} {510}},\ \bibinfo {pages} {351–355} (\bibinfo {year} {2014})}\BibitemShut {NoStop}%
\bibitem [{\citenamefont {Macedo}\ \emph {et~al.}(2025)\citenamefont {Macedo}, \citenamefont {Andriolo}, \citenamefont {Zamora}, \citenamefont {Poderini},\ and\ \citenamefont {Chaves}}]{Macedo_2025}%
  \BibitemOpen
  \bibfield  {author} {\bibinfo {author} {\bibfnamefont {R.~A.}\ \bibnamefont {Macedo}}, \bibinfo {author} {\bibfnamefont {P.}~\bibnamefont {Andriolo}}, \bibinfo {author} {\bibfnamefont {S.}~\bibnamefont {Zamora}}, \bibinfo {author} {\bibfnamefont {D.}~\bibnamefont {Poderini}},\ and\ \bibinfo {author} {\bibfnamefont {R.}~\bibnamefont {Chaves}},\ }\href {https://doi.org/10.48550/ARXIV.2503.18734} {\bibinfo {title} {Witnessing magic with bell inequalities}} (\bibinfo {year} {2025})\BibitemShut {NoStop}%
\bibitem [{\citenamefont {Cusumano}\ \emph {et~al.}(2025)\citenamefont {Cusumano}, \citenamefont {Venuti}, \citenamefont {Cepollaro}, \citenamefont {Esposito}, \citenamefont {Iannotti}, \citenamefont {Jasser}, \citenamefont {Odavi\'c}, \citenamefont {Viscardi},\ and\ \citenamefont {Hamma}}]{Cusumano_2025}%
  \BibitemOpen
  \bibfield  {author} {\bibinfo {author} {\bibfnamefont {S.}~\bibnamefont {Cusumano}}, \bibinfo {author} {\bibfnamefont {L.~C.}\ \bibnamefont {Venuti}}, \bibinfo {author} {\bibfnamefont {S.}~\bibnamefont {Cepollaro}}, \bibinfo {author} {\bibfnamefont {G.}~\bibnamefont {Esposito}}, \bibinfo {author} {\bibfnamefont {D.}~\bibnamefont {Iannotti}}, \bibinfo {author} {\bibfnamefont {B.}~\bibnamefont {Jasser}}, \bibinfo {author} {\bibfnamefont {J.}~\bibnamefont {Odavi\'c}}, \bibinfo {author} {\bibfnamefont {M.}~\bibnamefont {Viscardi}},\ and\ \bibinfo {author} {\bibfnamefont {A.}~\bibnamefont {Hamma}},\ }\href {https://arxiv.org/abs/2504.03351} {\bibinfo {title} {Non-stabilizerness and violations of chsh inequalities}} (\bibinfo {year} {2025}),\ \Eprint {https://arxiv.org/abs/2504.03351} {arXiv:2504.03351 [quant-ph]} \BibitemShut {NoStop}%
\bibitem [{\citenamefont {Horodecki}\ \emph {et~al.}(1995)\citenamefont {Horodecki}, \citenamefont {Horodecki},\ and\ \citenamefont {Horodecki}}]{Horodecki_1995}%
  \BibitemOpen
  \bibfield  {author} {\bibinfo {author} {\bibfnamefont {R.}~\bibnamefont {Horodecki}}, \bibinfo {author} {\bibfnamefont {P.}~\bibnamefont {Horodecki}},\ and\ \bibinfo {author} {\bibfnamefont {M.}~\bibnamefont {Horodecki}},\ }\href {https://doi.org/10.1016/0375-9601(95)00214-n} {\bibfield  {journal} {\bibinfo  {journal} {Physics Letters A}\ }\textbf {\bibinfo {volume} {200}},\ \bibinfo {pages} {340–344} (\bibinfo {year} {1995})}\BibitemShut {NoStop}%
\bibitem [{\citenamefont {Nagata}\ \emph {et~al.}(2004)\citenamefont {Nagata}, \citenamefont {Laskowski}, \citenamefont {Wie\ifmmode~\acute{s}\else \'{s}\fi{}niak},\ and\ \citenamefont {\ifmmode~\dot{Z}\else \.{Z}\fi{}ukowski}}]{Nagata_2004}%
  \BibitemOpen
  \bibfield  {author} {\bibinfo {author} {\bibfnamefont {K.}~\bibnamefont {Nagata}}, \bibinfo {author} {\bibfnamefont {W.}~\bibnamefont {Laskowski}}, \bibinfo {author} {\bibfnamefont {M.}~\bibnamefont {Wie\ifmmode~\acute{s}\else \'{s}\fi{}niak}},\ and\ \bibinfo {author} {\bibfnamefont {M.}~\bibnamefont {\ifmmode~\dot{Z}\else \.{Z}\fi{}ukowski}},\ }\href {https://doi.org/10.1103/PhysRevLett.93.230403} {\bibfield  {journal} {\bibinfo  {journal} {Phys. Rev. Lett.}\ }\textbf {\bibinfo {volume} {93}},\ \bibinfo {pages} {230403} (\bibinfo {year} {2004})}\BibitemShut {NoStop}%
\bibitem [{\citenamefont {Hassan}\ and\ \citenamefont {Joag}(2008)}]{Hassan_2008}%
  \BibitemOpen
  \bibfield  {author} {\bibinfo {author} {\bibfnamefont {A.~S.~M.}\ \bibnamefont {Hassan}}\ and\ \bibinfo {author} {\bibfnamefont {P.~S.}\ \bibnamefont {Joag}},\ }\href {https://doi.org/10.1103/PhysRevA.77.062334} {\bibfield  {journal} {\bibinfo  {journal} {Phys. Rev. A}\ }\textbf {\bibinfo {volume} {77}},\ \bibinfo {pages} {062334} (\bibinfo {year} {2008})}\BibitemShut {NoStop}%
\bibitem [{\citenamefont {Badzia\ifmmode~\mbox{\c{}}\else \c{}\fi{}g}\ \emph {et~al.}(2008)\citenamefont {Badzia\ifmmode~\mbox{\c{}}\else \c{}\fi{}g}, \citenamefont {Brukner}, \citenamefont {Laskowski}, \citenamefont {Paterek},\ and\ \citenamefont {\ifmmode~\dot{Z}\else \.{Z}\fi{}ukowski}}]{Badziag_2008}%
  \BibitemOpen
  \bibfield  {author} {\bibinfo {author} {\bibfnamefont {P.}~\bibnamefont {Badzia\ifmmode~\mbox{\c{}}\else \c{}\fi{}g}}, \bibinfo {author} {\bibfnamefont {{\v C}.}~\bibnamefont {Brukner}}, \bibinfo {author} {\bibfnamefont {W.}~\bibnamefont {Laskowski}}, \bibinfo {author} {\bibfnamefont {T.}~\bibnamefont {Paterek}},\ and\ \bibinfo {author} {\bibfnamefont {M.}~\bibnamefont {\ifmmode~\dot{Z}\else \.{Z}\fi{}ukowski}},\ }\href {https://doi.org/10.1103/PhysRevLett.100.140403} {\bibfield  {journal} {\bibinfo  {journal} {Phys. Rev. Lett.}\ }\textbf {\bibinfo {volume} {100}},\ \bibinfo {pages} {140403} (\bibinfo {year} {2008})}\BibitemShut {NoStop}%
\bibitem [{\citenamefont {Tran}\ \emph {et~al.}(2015)\citenamefont {Tran}, \citenamefont {Dakić}, \citenamefont {Arnault}, \citenamefont {Laskowski},\ and\ \citenamefont {Paterek}}]{Tran_2015}%
  \BibitemOpen
  \bibfield  {author} {\bibinfo {author} {\bibfnamefont {M.~C.}\ \bibnamefont {Tran}}, \bibinfo {author} {\bibfnamefont {B.}~\bibnamefont {Dakić}}, \bibinfo {author} {\bibfnamefont {F.}~\bibnamefont {Arnault}}, \bibinfo {author} {\bibfnamefont {W.}~\bibnamefont {Laskowski}},\ and\ \bibinfo {author} {\bibfnamefont {T.}~\bibnamefont {Paterek}},\ }\href {https://doi.org/10.1103/physreva.92.050301} {\bibfield  {journal} {\bibinfo  {journal} {Physical Review A}\ }\textbf {\bibinfo {volume} {92}},\ \bibinfo {pages} {050301} (\bibinfo {year} {2015})}\BibitemShut {NoStop}%
\bibitem [{\citenamefont {Ketterer}\ \emph {et~al.}(2019)\citenamefont {Ketterer}, \citenamefont {Wyderka},\ and\ \citenamefont {G\"{u}hne}}]{Ketterer_2019}%
  \BibitemOpen
  \bibfield  {author} {\bibinfo {author} {\bibfnamefont {A.}~\bibnamefont {Ketterer}}, \bibinfo {author} {\bibfnamefont {N.}~\bibnamefont {Wyderka}},\ and\ \bibinfo {author} {\bibfnamefont {O.}~\bibnamefont {G\"{u}hne}},\ }\href {https://doi.org/10.1103/physrevlett.122.120505} {\bibfield  {journal} {\bibinfo  {journal} {Physical Review Letters}\ }\textbf {\bibinfo {volume} {122}},\ \bibinfo {pages} {120505} (\bibinfo {year} {2019})}\BibitemShut {NoStop}%
\bibitem [{\citenamefont {Wyderka}\ and\ \citenamefont {G\"{u}hne}(2020)}]{Wyderka_2020}%
  \BibitemOpen
  \bibfield  {author} {\bibinfo {author} {\bibfnamefont {N.}~\bibnamefont {Wyderka}}\ and\ \bibinfo {author} {\bibfnamefont {O.}~\bibnamefont {G\"{u}hne}},\ }\href {https://doi.org/10.1088/1751-8121/ab7f0a} {\bibfield  {journal} {\bibinfo  {journal} {Journal of Physics A: Mathematical and Theoretical}\ }\textbf {\bibinfo {volume} {53}},\ \bibinfo {pages} {345302} (\bibinfo {year} {2020})}\BibitemShut {NoStop}%
\bibitem [{\citenamefont {Imai}\ \emph {et~al.}(2021)\citenamefont {Imai}, \citenamefont {Wyderka}, \citenamefont {Ketterer},\ and\ \citenamefont {G\"uhne}}]{Imai_2021}%
  \BibitemOpen
  \bibfield  {author} {\bibinfo {author} {\bibfnamefont {S.}~\bibnamefont {Imai}}, \bibinfo {author} {\bibfnamefont {N.}~\bibnamefont {Wyderka}}, \bibinfo {author} {\bibfnamefont {A.}~\bibnamefont {Ketterer}},\ and\ \bibinfo {author} {\bibfnamefont {O.}~\bibnamefont {G\"uhne}},\ }\href {https://doi.org/10.1103/PhysRevLett.126.150501} {\bibfield  {journal} {\bibinfo  {journal} {Phys. Rev. Lett.}\ }\textbf {\bibinfo {volume} {126}},\ \bibinfo {pages} {150501} (\bibinfo {year} {2021})}\BibitemShut {NoStop}%
\bibitem [{\citenamefont {Ketterer}\ \emph {et~al.}(2022)\citenamefont {Ketterer}, \citenamefont {Imai}, \citenamefont {Wyderka},\ and\ \citenamefont {G\"uhne}}]{Ketterer_2022}%
  \BibitemOpen
  \bibfield  {author} {\bibinfo {author} {\bibfnamefont {A.}~\bibnamefont {Ketterer}}, \bibinfo {author} {\bibfnamefont {S.}~\bibnamefont {Imai}}, \bibinfo {author} {\bibfnamefont {N.}~\bibnamefont {Wyderka}},\ and\ \bibinfo {author} {\bibfnamefont {O.}~\bibnamefont {G\"uhne}},\ }\href {https://doi.org/10.1103/PhysRevA.106.L010402} {\bibfield  {journal} {\bibinfo  {journal} {Phys. Rev. A}\ }\textbf {\bibinfo {volume} {106}},\ \bibinfo {pages} {L010402} (\bibinfo {year} {2022})}\BibitemShut {NoStop}%
\bibitem [{\citenamefont {Cieśliński}\ \emph {et~al.}(2024)\citenamefont {Cieśliński}, \citenamefont {Knips}, \citenamefont {Kowalczyk}, \citenamefont {Laskowski}, \citenamefont {Paterek}, \citenamefont {Vértesi},\ and\ \citenamefont {Weinfurter}}]{Cieslinski_2024poly}%
  \BibitemOpen
  \bibfield  {author} {\bibinfo {author} {\bibfnamefont {P.}~\bibnamefont {Cieśliński}}, \bibinfo {author} {\bibfnamefont {L.}~\bibnamefont {Knips}}, \bibinfo {author} {\bibfnamefont {M.}~\bibnamefont {Kowalczyk}}, \bibinfo {author} {\bibfnamefont {W.}~\bibnamefont {Laskowski}}, \bibinfo {author} {\bibfnamefont {T.}~\bibnamefont {Paterek}}, \bibinfo {author} {\bibfnamefont {T.}~\bibnamefont {Vértesi}},\ and\ \bibinfo {author} {\bibfnamefont {H.}~\bibnamefont {Weinfurter}},\ }\href {https://doi.org/10.1073/pnas.2404455121} {\bibfield  {journal} {\bibinfo  {journal} {Proceedings of the National Academy of Sciences}\ }\textbf {\bibinfo {volume} {121}},\ \bibinfo {pages} {e2404455121} (\bibinfo {year} {2024})}\BibitemShut {NoStop}%
\bibitem [{\citenamefont {D\"ur}\ \emph {et~al.}(2000)\citenamefont {D\"ur}, \citenamefont {Vidal},\ and\ \citenamefont {Cirac}}]{Dur_2000}%
  \BibitemOpen
  \bibfield  {author} {\bibinfo {author} {\bibfnamefont {W.}~\bibnamefont {D\"ur}}, \bibinfo {author} {\bibfnamefont {G.}~\bibnamefont {Vidal}},\ and\ \bibinfo {author} {\bibfnamefont {J.~I.}\ \bibnamefont {Cirac}},\ }\href {https://doi.org/10.1103/PhysRevA.62.062314} {\bibfield  {journal} {\bibinfo  {journal} {Phys. Rev. A}\ }\textbf {\bibinfo {volume} {62}},\ \bibinfo {pages} {062314} (\bibinfo {year} {2000})}\BibitemShut {NoStop}%
\bibitem [{\citenamefont {Gruca}\ \emph {et~al.}(2010)\citenamefont {Gruca}, \citenamefont {Laskowski}, \citenamefont {\ifmmode~\dot{Z}\else \.{Z}\fi{}ukowski}, \citenamefont {Kiesel}, \citenamefont {Wieczorek}, \citenamefont {Schmid},\ and\ \citenamefont {Weinfurter}}]{Gruca_2010}%
  \BibitemOpen
  \bibfield  {author} {\bibinfo {author} {\bibfnamefont {J.}~\bibnamefont {Gruca}}, \bibinfo {author} {\bibfnamefont {W.}~\bibnamefont {Laskowski}}, \bibinfo {author} {\bibfnamefont {M.}~\bibnamefont {\ifmmode~\dot{Z}\else \.{Z}\fi{}ukowski}}, \bibinfo {author} {\bibfnamefont {N.}~\bibnamefont {Kiesel}}, \bibinfo {author} {\bibfnamefont {W.}~\bibnamefont {Wieczorek}}, \bibinfo {author} {\bibfnamefont {C.}~\bibnamefont {Schmid}},\ and\ \bibinfo {author} {\bibfnamefont {H.}~\bibnamefont {Weinfurter}},\ }\href {https://doi.org/10.1103/PhysRevA.82.012118} {\bibfield  {journal} {\bibinfo  {journal} {Phys. Rev. A}\ }\textbf {\bibinfo {volume} {82}},\ \bibinfo {pages} {012118} (\bibinfo {year} {2010})}\BibitemShut {NoStop}%
\bibitem [{\citenamefont {Wieczorek}\ \emph {et~al.}(2008)\citenamefont {Wieczorek}, \citenamefont {Schmid}, \citenamefont {Kiesel}, \citenamefont {Pohlner}, \citenamefont {G\"uhne},\ and\ \citenamefont {Weinfurter}}]{Wieczorek_2008}%
  \BibitemOpen
  \bibfield  {author} {\bibinfo {author} {\bibfnamefont {W.}~\bibnamefont {Wieczorek}}, \bibinfo {author} {\bibfnamefont {C.}~\bibnamefont {Schmid}}, \bibinfo {author} {\bibfnamefont {N.}~\bibnamefont {Kiesel}}, \bibinfo {author} {\bibfnamefont {R.}~\bibnamefont {Pohlner}}, \bibinfo {author} {\bibfnamefont {O.}~\bibnamefont {G\"uhne}},\ and\ \bibinfo {author} {\bibfnamefont {H.}~\bibnamefont {Weinfurter}},\ }\href {https://doi.org/10.1103/PhysRevLett.101.010503} {\bibfield  {journal} {\bibinfo  {journal} {Phys. Rev. Lett.}\ }\textbf {\bibinfo {volume} {101}},\ \bibinfo {pages} {010503} (\bibinfo {year} {2008})}\BibitemShut {NoStop}%
\bibitem [{\citenamefont {Schmid}(2008)}]{Schmid_2008phd}%
  \BibitemOpen
  \bibfield  {author} {\bibinfo {author} {\bibfnamefont {C.}~\bibnamefont {Schmid}},\ }\emph {\bibinfo {title} {Multi-photon entanglement and applications in quantum information}},\ \href {https://doi.org/10.5282/EDOC.8847} {Ph.D. thesis},\ \bibinfo  {school} {Ludwig-Maximilians-Universit\"{a}t M\"{u}nchen} (\bibinfo {year} {2008})\BibitemShut {NoStop}%
\bibitem [{\citenamefont {Weinfurter}\ and\ \citenamefont {\ifmmode~\dot{Z}\else \.{Z}\fi{}ukowski}(2001)}]{Weinfurter_2001}%
  \BibitemOpen
  \bibfield  {author} {\bibinfo {author} {\bibfnamefont {H.}~\bibnamefont {Weinfurter}}\ and\ \bibinfo {author} {\bibfnamefont {M.}~\bibnamefont {\ifmmode~\dot{Z}\else \.{Z}\fi{}ukowski}},\ }\href {https://doi.org/10.1103/PhysRevA.64.010102} {\bibfield  {journal} {\bibinfo  {journal} {Phys. Rev. A}\ }\textbf {\bibinfo {volume} {64}},\ \bibinfo {pages} {010102} (\bibinfo {year} {2001})}\BibitemShut {NoStop}%
\bibitem [{\citenamefont {Eibl}\ \emph {et~al.}(2003)\citenamefont {Eibl}, \citenamefont {Gaertner}, \citenamefont {Bourennane}, \citenamefont {Kurtsiefer}, \citenamefont {\ifmmode~\dot{Z}\else \.{Z}\fi{}ukowski},\ and\ \citenamefont {Weinfurter}}]{Eibl_2003}%
  \BibitemOpen
  \bibfield  {author} {\bibinfo {author} {\bibfnamefont {M.}~\bibnamefont {Eibl}}, \bibinfo {author} {\bibfnamefont {S.}~\bibnamefont {Gaertner}}, \bibinfo {author} {\bibfnamefont {M.}~\bibnamefont {Bourennane}}, \bibinfo {author} {\bibfnamefont {C.}~\bibnamefont {Kurtsiefer}}, \bibinfo {author} {\bibfnamefont {M.}~\bibnamefont {\ifmmode~\dot{Z}\else \.{Z}\fi{}ukowski}},\ and\ \bibinfo {author} {\bibfnamefont {H.}~\bibnamefont {Weinfurter}},\ }\href {https://doi.org/10.1103/PhysRevLett.90.200403} {\bibfield  {journal} {\bibinfo  {journal} {Phys. Rev. Lett.}\ }\textbf {\bibinfo {volume} {90}},\ \bibinfo {pages} {200403} (\bibinfo {year} {2003})}\BibitemShut {NoStop}%
\bibitem [{\citenamefont {Knips}\ \emph {et~al.}(2016)\citenamefont {Knips}, \citenamefont {Schwemmer}, \citenamefont {Klein}, \citenamefont {Wie\ifmmode~\acute{s}\else \'{s}\fi{}niak},\ and\ \citenamefont {Weinfurter}}]{PhysRevLett.117.210504}%
  \BibitemOpen
  \bibfield  {author} {\bibinfo {author} {\bibfnamefont {L.}~\bibnamefont {Knips}}, \bibinfo {author} {\bibfnamefont {C.}~\bibnamefont {Schwemmer}}, \bibinfo {author} {\bibfnamefont {N.}~\bibnamefont {Klein}}, \bibinfo {author} {\bibfnamefont {M.}~\bibnamefont {Wie\ifmmode~\acute{s}\else \'{s}\fi{}niak}},\ and\ \bibinfo {author} {\bibfnamefont {H.}~\bibnamefont {Weinfurter}},\ }\href {https://doi.org/10.1103/PhysRevLett.117.210504} {\bibfield  {journal} {\bibinfo  {journal} {Phys. Rev. Lett.}\ }\textbf {\bibinfo {volume} {117}},\ \bibinfo {pages} {210504} (\bibinfo {year} {2016})}\BibitemShut {NoStop}%
\bibitem [{\citenamefont {Krischek}\ \emph {et~al.}(2011)\citenamefont {Krischek}, \citenamefont {Schwemmer}, \citenamefont {Wieczorek}, \citenamefont {Weinfurter}, \citenamefont {Hyllus}, \citenamefont {Pezz\'e},\ and\ \citenamefont {Smerzi}}]{PhysRevLett.107.080504}%
  \BibitemOpen
  \bibfield  {author} {\bibinfo {author} {\bibfnamefont {R.}~\bibnamefont {Krischek}}, \bibinfo {author} {\bibfnamefont {C.}~\bibnamefont {Schwemmer}}, \bibinfo {author} {\bibfnamefont {W.}~\bibnamefont {Wieczorek}}, \bibinfo {author} {\bibfnamefont {H.}~\bibnamefont {Weinfurter}}, \bibinfo {author} {\bibfnamefont {P.}~\bibnamefont {Hyllus}}, \bibinfo {author} {\bibfnamefont {L.}~\bibnamefont {Pezz\'e}},\ and\ \bibinfo {author} {\bibfnamefont {A.}~\bibnamefont {Smerzi}},\ }\href {https://doi.org/10.1103/PhysRevLett.107.080504} {\bibfield  {journal} {\bibinfo  {journal} {Phys. Rev. Lett.}\ }\textbf {\bibinfo {volume} {107}},\ \bibinfo {pages} {080504} (\bibinfo {year} {2011})}\BibitemShut {NoStop}%
\bibitem [{\citenamefont {Schmid}\ \emph {et~al.}(2010)\citenamefont {Schmid}, \citenamefont {Flitney}, \citenamefont {Wieczorek}, \citenamefont {Kiesel}, \citenamefont {Weinfurter},\ and\ \citenamefont {Hollenberg}}]{Schmid_2010}%
  \BibitemOpen
  \bibfield  {author} {\bibinfo {author} {\bibfnamefont {C.}~\bibnamefont {Schmid}}, \bibinfo {author} {\bibfnamefont {A.~P.}\ \bibnamefont {Flitney}}, \bibinfo {author} {\bibfnamefont {W.}~\bibnamefont {Wieczorek}}, \bibinfo {author} {\bibfnamefont {N.}~\bibnamefont {Kiesel}}, \bibinfo {author} {\bibfnamefont {H.}~\bibnamefont {Weinfurter}},\ and\ \bibinfo {author} {\bibfnamefont {L.~C.~L.}\ \bibnamefont {Hollenberg}},\ }\href {https://doi.org/10.1088/1367-2630/12/6/063031} {\bibfield  {journal} {\bibinfo  {journal} {New Journal of Physics}\ }\textbf {\bibinfo {volume} {12}},\ \bibinfo {pages} {063031} (\bibinfo {year} {2010})}\BibitemShut {NoStop}%
\bibitem [{\citenamefont {Bravyi}\ and\ \citenamefont {Kitaev}(2005)}]{Bravyi_2005}%
  \BibitemOpen
  \bibfield  {author} {\bibinfo {author} {\bibfnamefont {S.}~\bibnamefont {Bravyi}}\ and\ \bibinfo {author} {\bibfnamefont {A.}~\bibnamefont {Kitaev}},\ }\href {https://doi.org/10.1103/PhysRevA.71.022316} {\bibfield  {journal} {\bibinfo  {journal} {Phys. Rev. A}\ }\textbf {\bibinfo {volume} {71}},\ \bibinfo {pages} {022316} (\bibinfo {year} {2005})}\BibitemShut {NoStop}%
\bibitem [{\citenamefont {Bravyi}\ and\ \citenamefont {Gosset}(2016)}]{Bravyi_2016}%
  \BibitemOpen
  \bibfield  {author} {\bibinfo {author} {\bibfnamefont {S.}~\bibnamefont {Bravyi}}\ and\ \bibinfo {author} {\bibfnamefont {D.}~\bibnamefont {Gosset}},\ }\href {https://doi.org/10.1103/PhysRevLett.116.250501} {\bibfield  {journal} {\bibinfo  {journal} {Phys. Rev. Lett.}\ }\textbf {\bibinfo {volume} {116}},\ \bibinfo {pages} {250501} (\bibinfo {year} {2016})}\BibitemShut {NoStop}%
\bibitem [{\citenamefont {Howard}\ and\ \citenamefont {Campbell}(2017)}]{Howard_2017}%
  \BibitemOpen
  \bibfield  {author} {\bibinfo {author} {\bibfnamefont {M.}~\bibnamefont {Howard}}\ and\ \bibinfo {author} {\bibfnamefont {E.}~\bibnamefont {Campbell}},\ }\href {https://doi.org/10.1103/PhysRevLett.118.090501} {\bibfield  {journal} {\bibinfo  {journal} {Phys. Rev. Lett.}\ }\textbf {\bibinfo {volume} {118}},\ \bibinfo {pages} {090501} (\bibinfo {year} {2017})}\BibitemShut {NoStop}%
\bibitem [{\citenamefont {Liu}\ and\ \citenamefont {Winter}(2022)}]{Liu_2022}%
  \BibitemOpen
  \bibfield  {author} {\bibinfo {author} {\bibfnamefont {Z.-W.}\ \bibnamefont {Liu}}\ and\ \bibinfo {author} {\bibfnamefont {A.}~\bibnamefont {Winter}},\ }\href {https://doi.org/10.1103/PRXQuantum.3.020333} {\bibfield  {journal} {\bibinfo  {journal} {PRX Quantum}\ }\textbf {\bibinfo {volume} {3}},\ \bibinfo {pages} {020333} (\bibinfo {year} {2022})}\BibitemShut {NoStop}%
\bibitem [{\citenamefont {Oliviero}\ \emph {et~al.}(2022)\citenamefont {Oliviero}, \citenamefont {Leone}, \citenamefont {Zhou},\ and\ \citenamefont {Hamma}}]{Salvatore_2022}%
  \BibitemOpen
  \bibfield  {author} {\bibinfo {author} {\bibfnamefont {S.~F.~E.}\ \bibnamefont {Oliviero}}, \bibinfo {author} {\bibfnamefont {L.}~\bibnamefont {Leone}}, \bibinfo {author} {\bibfnamefont {Y.}~\bibnamefont {Zhou}},\ and\ \bibinfo {author} {\bibfnamefont {A.}~\bibnamefont {Hamma}},\ }\href {https://doi.org/10.21468/SciPostPhys.12.3.096} {\bibfield  {journal} {\bibinfo  {journal} {SciPost Phys.}\ }\textbf {\bibinfo {volume} {12}},\ \bibinfo {pages} {096} (\bibinfo {year} {2022})}\BibitemShut {NoStop}%
\bibitem [{\citenamefont {White}\ and\ \citenamefont {White}(2024)}]{White_2024}%
  \BibitemOpen
  \bibfield  {author} {\bibinfo {author} {\bibfnamefont {C.~D.}\ \bibnamefont {White}}\ and\ \bibinfo {author} {\bibfnamefont {M.~J.}\ \bibnamefont {White}},\ }\href {https://doi.org/10.1103/PhysRevD.110.116016} {\bibfield  {journal} {\bibinfo  {journal} {Phys. Rev. D}\ }\textbf {\bibinfo {volume} {110}},\ \bibinfo {pages} {116016} (\bibinfo {year} {2024})}\BibitemShut {NoStop}%
\bibitem [{\citenamefont {Tirrito}\ \emph {et~al.}(2024)\citenamefont {Tirrito}, \citenamefont {Tarabunga}, \citenamefont {Lami}, \citenamefont {Chanda}, \citenamefont {Leone}, \citenamefont {Oliviero}, \citenamefont {Dalmonte}, \citenamefont {Collura},\ and\ \citenamefont {Hamma}}]{Tirrito_2024}%
  \BibitemOpen
  \bibfield  {author} {\bibinfo {author} {\bibfnamefont {E.}~\bibnamefont {Tirrito}}, \bibinfo {author} {\bibfnamefont {P.~S.}\ \bibnamefont {Tarabunga}}, \bibinfo {author} {\bibfnamefont {G.}~\bibnamefont {Lami}}, \bibinfo {author} {\bibfnamefont {T.}~\bibnamefont {Chanda}}, \bibinfo {author} {\bibfnamefont {L.}~\bibnamefont {Leone}}, \bibinfo {author} {\bibfnamefont {S.~F.~E.}\ \bibnamefont {Oliviero}}, \bibinfo {author} {\bibfnamefont {M.}~\bibnamefont {Dalmonte}}, \bibinfo {author} {\bibfnamefont {M.}~\bibnamefont {Collura}},\ and\ \bibinfo {author} {\bibfnamefont {A.}~\bibnamefont {Hamma}},\ }\href {https://doi.org/10.1103/PhysRevA.109.L040401} {\bibfield  {journal} {\bibinfo  {journal} {Phys. Rev. A}\ }\textbf {\bibinfo {volume} {109}},\ \bibinfo {pages} {L040401} (\bibinfo {year} {2024})}\BibitemShut {NoStop}%
\bibitem [{\citenamefont {Turkeshi}\ \emph {et~al.}(2025)\citenamefont {Turkeshi}, \citenamefont {Dymarsky},\ and\ \citenamefont {Sierant}}]{Turkeshi_2025}%
  \BibitemOpen
  \bibfield  {author} {\bibinfo {author} {\bibfnamefont {X.}~\bibnamefont {Turkeshi}}, \bibinfo {author} {\bibfnamefont {A.}~\bibnamefont {Dymarsky}},\ and\ \bibinfo {author} {\bibfnamefont {P.}~\bibnamefont {Sierant}},\ }\href {https://doi.org/10.1103/PhysRevB.111.054301} {\bibfield  {journal} {\bibinfo  {journal} {Phys. Rev. B}\ }\textbf {\bibinfo {volume} {111}},\ \bibinfo {pages} {054301} (\bibinfo {year} {2025})}\BibitemShut {NoStop}%
\bibitem [{\citenamefont {Br\"okemeier}\ \emph {et~al.}(2025)\citenamefont {Br\"okemeier}, \citenamefont {Hengstenberg}, \citenamefont {Keeble}, \citenamefont {Robin}, \citenamefont {Rocco},\ and\ \citenamefont {Savage}}]{Brokemeier_2025}%
  \BibitemOpen
  \bibfield  {author} {\bibinfo {author} {\bibfnamefont {F.}~\bibnamefont {Br\"okemeier}}, \bibinfo {author} {\bibfnamefont {S.~M.}\ \bibnamefont {Hengstenberg}}, \bibinfo {author} {\bibfnamefont {J.~W.~T.}\ \bibnamefont {Keeble}}, \bibinfo {author} {\bibfnamefont {C.~E.~P.}\ \bibnamefont {Robin}}, \bibinfo {author} {\bibfnamefont {F.}~\bibnamefont {Rocco}},\ and\ \bibinfo {author} {\bibfnamefont {M.~J.}\ \bibnamefont {Savage}},\ }\href {https://doi.org/10.1103/PhysRevC.111.034317} {\bibfield  {journal} {\bibinfo  {journal} {Phys. Rev. C}\ }\textbf {\bibinfo {volume} {111}},\ \bibinfo {pages} {034317} (\bibinfo {year} {2025})}\BibitemShut {NoStop}%
\bibitem [{\citenamefont {Hoshino}\ \emph {et~al.}(2025)\citenamefont {Hoshino}, \citenamefont {Oshikawa},\ and\ \citenamefont {Ashida}}]{Hoshino_2025}%
  \BibitemOpen
  \bibfield  {author} {\bibinfo {author} {\bibfnamefont {M.}~\bibnamefont {Hoshino}}, \bibinfo {author} {\bibfnamefont {M.}~\bibnamefont {Oshikawa}},\ and\ \bibinfo {author} {\bibfnamefont {Y.}~\bibnamefont {Ashida}},\ }\href {https://doi.org/10.48550/ARXIV.2503.13599} {\bibinfo {title} {Stabilizer rényi entropy and conformal field theory}} (\bibinfo {year} {2025})\BibitemShut {NoStop}%
\bibitem [{\citenamefont {Odavi\ifmmode~\acute{c}\else \'{c}\fi{}}\ \emph {et~al.}(2025)\citenamefont {Odavi\ifmmode~\acute{c}\else \'{c}\fi{}}, \citenamefont {Viscardi},\ and\ \citenamefont {Hamma}}]{Odavi_2025}%
  \BibitemOpen
  \bibfield  {author} {\bibinfo {author} {\bibfnamefont {J.}~\bibnamefont {Odavi\ifmmode~\acute{c}\else \'{c}\fi{}}}, \bibinfo {author} {\bibfnamefont {M.}~\bibnamefont {Viscardi}},\ and\ \bibinfo {author} {\bibfnamefont {A.}~\bibnamefont {Hamma}},\ }\href {https://doi.org/10.1103/y9r6-dx7p} {\bibfield  {journal} {\bibinfo  {journal} {Phys. Rev. B}\ }\textbf {\bibinfo {volume} {112}},\ \bibinfo {pages} {104301} (\bibinfo {year} {2025})}\BibitemShut {NoStop}%
\bibitem [{\citenamefont {Hou}\ \emph {et~al.}(2025)\citenamefont {Hou}, \citenamefont {Cao},\ and\ \citenamefont {Yang}}]{Hou_2025}%
  \BibitemOpen
  \bibfield  {author} {\bibinfo {author} {\bibfnamefont {Z.-Y.}\ \bibnamefont {Hou}}, \bibinfo {author} {\bibfnamefont {C.}~\bibnamefont {Cao}},\ and\ \bibinfo {author} {\bibfnamefont {Z.-C.}\ \bibnamefont {Yang}},\ }\href {https://doi.org/10.48550/ARXIV.2503.20873} {\bibinfo {title} {Stabilizer entanglement enhances magic injection}} (\bibinfo {year} {2025})\BibitemShut {NoStop}%
\bibitem [{\citenamefont {Catalano}\ \emph {et~al.}(2024)\citenamefont {Catalano}, \citenamefont {Odavić}, \citenamefont {Torre}, \citenamefont {Hamma}, \citenamefont {Franchini},\ and\ \citenamefont {Giampaolo}}]{Catalano_2025}%
  \BibitemOpen
  \bibfield  {author} {\bibinfo {author} {\bibfnamefont {A.~G.}\ \bibnamefont {Catalano}}, \bibinfo {author} {\bibfnamefont {J.}~\bibnamefont {Odavić}}, \bibinfo {author} {\bibfnamefont {G.}~\bibnamefont {Torre}}, \bibinfo {author} {\bibfnamefont {A.}~\bibnamefont {Hamma}}, \bibinfo {author} {\bibfnamefont {F.}~\bibnamefont {Franchini}},\ and\ \bibinfo {author} {\bibfnamefont {S.~M.}\ \bibnamefont {Giampaolo}},\ }\href {https://doi.org/10.48550/ARXIV.2406.19457} {\bibinfo {title} {Magic phase transition and non-local complexity in generalized $w$ state}} (\bibinfo {year} {2024})\BibitemShut {NoStop}%
\bibitem [{\citenamefont {Junior}\ \emph {et~al.}(2025)\citenamefont {Junior}, \citenamefont {Zamora}, \citenamefont {Mac\^edo}, \citenamefont {Sarubi}, \citenamefont {Varela}, \citenamefont {Rocha}, \citenamefont {Moreira},\ and\ \citenamefont {Chaves}}]{Junior_2025}%
  \BibitemOpen
  \bibfield  {author} {\bibinfo {author} {\bibfnamefont {A.~B.~P.}\ \bibnamefont {Junior}}, \bibinfo {author} {\bibfnamefont {S.}~\bibnamefont {Zamora}}, \bibinfo {author} {\bibfnamefont {R.~A.}\ \bibnamefont {Mac\^edo}}, \bibinfo {author} {\bibfnamefont {T.~S.}\ \bibnamefont {Sarubi}}, \bibinfo {author} {\bibfnamefont {J.~M.}\ \bibnamefont {Varela}}, \bibinfo {author} {\bibfnamefont {G.~W.~C.}\ \bibnamefont {Rocha}}, \bibinfo {author} {\bibfnamefont {D.~A.}\ \bibnamefont {Moreira}},\ and\ \bibinfo {author} {\bibfnamefont {R.}~\bibnamefont {Chaves}},\ }\href {https://doi.org/10.48550/ARXIV.2504.12518} {\bibinfo {title} {Geometric analysis of the stabilizer polytope for few-qubit systems}} (\bibinfo {year} {2025})\BibitemShut {NoStop}%
\bibitem [{\citenamefont {Hoggar}(1998)}]{Hoggar_1998}%
  \BibitemOpen
  \bibfield  {author} {\bibinfo {author} {\bibfnamefont {S.~G.}\ \bibnamefont {Hoggar}},\ }\href {https://doi.org/10.1023/a:1005009727232} {\bibfield  {journal} {\bibinfo  {journal} {Geometriae Dedicata}\ }\textbf {\bibinfo {volume} {69}},\ \bibinfo {pages} {287–289} (\bibinfo {year} {1998})}\BibitemShut {NoStop}%
\end{thebibliography}%

\end{document}